\documentclass[11pt]{article}
\usepackage{tgpagella}
\usepackage[T1]{fontenc}
\usepackage[utf8]{inputenc}
\pdfoutput=1

\usepackage{booktabs} %
\usepackage{fullpage}
\usepackage{xfrac}
\usepackage{algorithm}
\usepackage[noend]{algorithmic}
\usepackage{amsmath,amsthm,amsfonts}
\usepackage{amsmath}
\usepackage{amssymb}
\usepackage{mathrsfs}
\usepackage{verbatim}
\usepackage{enumitem}
\usepackage{bm}
\usepackage[pagebackref]{hyperref}
\usepackage[capitalise,noabbrev]{cleveref}
\usepackage{xr}
\usepackage{tikz}
\usepackage{subcaption}
\usepackage{float}
\usepackage{listings}
\usepackage{array}
\usepackage{makecell}

\usepackage[suppress]{color-edits}
\usepackage[]{color-edits}
\addauthor{ll}{blue}

\definecolor{DarkBlue}{rgb}{0.1,0.1,0.5}
\hypersetup{
	colorlinks=true,       %
	linkcolor=DarkBlue,          %
	citecolor=DarkBlue,        %
	filecolor=DarkBlue,      %
	urlcolor=DarkBlue,          %
	pdftitle={},
	pdfauthor={},
}

\usepackage{tabularx}
\usepackage{float,subcaption,placeins}
\usepackage{amsmath}
\usepackage{amsthm}
\newtheorem{definition}{Definition}[section]
\newtheorem{proposition}{Proposition}[section]
\newtheorem{assumption}{Assumption}[section]

\usepackage{thm-restate}
\usepackage{float}
\usepackage{enumitem}
\usepackage{xr}
\usepackage{graphicx}
\graphicspath{{./figures/}}
\usepackage{thmtools}
\usepackage{url}
\usepackage{bm}
\usepackage[normalem]{ulem}
\usepackage{balance}
\usepackage{lipsum}
\usepackage{etoc}
\usepackage{multirow}

\usepackage{comment}
\newcommand{\N}[0]{\epsilon}

\usepackage[numbers,sort&compress]{natbib}

\renewcommand\cite[1]{\citep{#1}}

\usepackage{pgfplots}

\pgfmathdeclarefunction{gauss}{2}{%
  \pgfmathparse{1/(#2*sqrt(2*pi))*exp(-((x-#1)^2)/(2*#2^2))}%
}

\usetikzlibrary{patterns}
\usetikzlibrary{arrows.meta}

\pgfplotsset{compat=1.10}
\usepgfplotslibrary{fillbetween}

\usepackage{chngcntr}
\usepackage{apptools}

\title{Discretion in the Loop: 

Human Expertise in Algorithm-Assisted College Advising}

\author{Kara Schechtman~\footnotemark[3]~ \quad \quad Benjamin Brandon\footnotemark[6]   \quad \quad  Jenise Stafford\footnotemark[6] \\\\ Hannah Li~\footnotemark[4]~~\thanks{H.L. and L.T.L. contributed equally to this work as co-principal investigators.}  \quad \quad Lydia T. Liu~\footnotemark[3]~~\footnotemark[2]\\
	\\
	\footnotemark[3]~~Department of Computer Science, Princeton University \\
			\footnotemark[6]~
            Georgia State University\\
			\footnotemark[4]~~Columbia Business School \\
}
\date{\today}

\begin{document}

\maketitle

\begin{abstract}

In higher education, many institutions use algorithmic alerts to flag at-risk students and deliver advising at scale. %
While much research has focused on evaluating algorithmic predictions, relatively little is known about how discretionary interventions by human experts shape outcomes in algorithm-assisted settings. We study this question using rich quantitative and qualitative data from a randomized controlled trial of an algorithm-assisted advising program at Georgia State University. Taking a mixed-methods approach, we examine whether and how advisors use context unavailable to an algorithm to guide interventions and influence student success. We develop a causal graphical framework for human expertise in the interventional setting, extending prior work on discretion in purely predictive settings. We then test a necessary condition for discretionary expertise using structured advisor logs and student outcomes data, identifying several interventions that meet the criterion for statistical significance. Accordingly, we estimate that 2 out of 3 interventions taken by advisors in the treatment arm were plausibly ``expertly targeted'' to students using non-algorithmic context.  Systematic qualitative analysis of advisor notes corroborates these findings, showing a pattern of advisors incorporating diverse forms of contextual information---such as personal circumstances, financial issues, and student engagement---into their decisions. Finally, we document heterogeneity in advising styles, finding that one style elicits more holistic information about students and is associated with improved graduation rates. Our results offer theoretical and practical insight into the real-world effectiveness of algorithm-supported college advising, and underscore the importance of accounting for human expertise in the design, evaluation, and implementation of algorithmic decision systems.
\end{abstract}

\section{Introduction}

As algorithm–assisted decision making becomes increasingly common in high-stakes public domains, there is growing recognition that evaluating these systems requires more than assessing model performance in isolation. In settings where algorithms inform---but do not replace---human decision makers, understanding how human experts actually use algorithmic tools is essential to assessing their real-world impact and equity implications \cite{greenDisparateInteractionsAlgorithmintheLoop2019,imaiExperimentalEvaluationAlgorithmassisted2023,rajiEvaluatingPredictionbasedInterventions2025,costonValidityPerspectiveEvaluating2023, mclaughlinFairnessMachineAssistedHuman2022, fazelpourDiscipliningDeliberationSociotechnical2025}. For example, in college advising, an algorithm may identify students at risk of dropping out, but it is up to a human counselor to have a conversation with the student and suggest a course of action that is tailored to the student’s specific circumstances. Despite the widespread deployment of such systems, empirical research remains limited on how human decision makers interpret algorithmic recommendations, exercise discretion, and influence downstream outcomes.

The Monitoring Advising Analytics to Promote Success (MAAPS) experiment at Georgia State University provides a rare opportunity to investigate these questions in a real-world setting. This multi-year (2016-2022) randomized controlled trial evaluated an algorithm-assisted advising program that offered ``intensive, proactive, technology-enhanced advisement and degree planning'' \cite{rossmanMAAPSAdvisingExperiment2023}, identifying students to support through both advising meetings and algorithmic early warning systems. Notably, the MAAPS program was evaluated via a carefully designed and well-documented experiment that demonstrated positive treatment effects on graduation and persistence \cite{rossmanMAAPSAdvisingExperiment2021,rossmanMAAPSAdvisingExperiment2023}.

While studies like the MAAPS experiment offer encouraging evidence that algorithmic early warning systems can improve student outcomes, they leave open important questions about why these interventions work when they do. A recent review, for instance, notes that existing studies offer ``largely speculative interpretations of intervention mechanisms and effects'' \cite{larrabee_sonderlund_efficacy_2019}. One key question is about the role of human discretion. When algorithms are embedded in human-in-the-loop decision processes, it becomes critical to ask: \emph{How much of the decision-making success (i.e., improvement in outcomes) is attributable to the algorithm versus the human expert? } %

Recognizing that human discretion could shape the impact of algorithm-assisted decision making, we hone in on the following research question:  \textit{What is the evidence for advisor discretion in algorithm-assisted advising, and what is its role in improving student outcomes?} Understanding the balance between algorithmic and human contributions is key to scaling algorithm-assisted advising, a strategy that is increasingly used by public universities to boost student success and close equity gaps \cite{renick2020predictive, feathers2023takeaways}. While algorithms can be scaled without capacity constraints, the human component cannot---reflecting a hard-won lesson of the social sciences, that interventions performing well in initial studies may not scale due to changes in the implementation, available resources, or personnel training \cite{jepsen2009class, list2022voltage, boutilier2024randomized}. If human discretion is a key driver of impact, scaling such systems will require not only algorithmic technologies but also investment in human expertise and skills.

To examine the role of human discretion in the success of algorithm-assisted advising, we analyze multifaceted data on advisor decision making within the treatment group of the MAAPS trial. The dataset includes administrative student data that forms the basis for algorithmic alerts, as well as structured, time-stamped records of advisor interactions and student outcomes for each semester (described more thoroughly in \S\ref{sec:data_methods:data}). The advisor interaction records include systematic documentation of the interventions they delivered during meetings\footnote{Interventions were selected from a list of 20 (see SI \ref{appendix:dataset:meeting_logs}).} with students and qualitative comments on their conversations. 

This mixture of qualitative and quantitative data provides a unique window into human decision making alongside algorithmic tools. Quantitative data alone often fail to capture how practitioners interpret and act on algorithmic outputs, and can abstract away the social context and judgment that shape real-world decisions \cite{selbst2019abstraction}. Qualitative data, like advisors' meeting notes, can provide insight into ``street-level'' discretionary work \cite{saxena2022unpacking,saxenaRethinkingRiskAlgorithmic2023,alkhatib_street-level_2019} that is not reflected by structured data. Such data are rarely collected in tandem with quantitative measures, especially in education.

Using a novel mixed-methods approach combining theoretical modeling, statistical testing, and qualitative analysis, we investigate one important form of human discretion, how MAAPS advisors use ``non-algorithmic information''---information unavailable to an algorithm---to target support to students. Our study makes two primary contributions: 

\paragraph{Contribution 1: Theoretical Model of Expertise.}

We develop a causal graphical framework to define and audit for human expertise in the interventional setting (\S\ref{sec:theory}), adding to a growing body of research studying information asymmetries in human-AI collaboration \cite{balakrishnanHumanAlgorithmCollaboration2025,guo2025value,holsteinConceptualFrameworkHuman2020,alurAuditingHumanExpertise2023,alurHumanExpertiseAlgorithmic2024,agarwalCombiningHumanExpertise2023,hoongImprovingAIAssistedDecisionMaking2025}. Extending prior work on auditing for human expertise in pure prediction tasks \cite{alurAuditingHumanExpertise2023,alurHumanExpertiseAlgorithmic2024}, we define \emph{expertly targeted action} in terms of three causal criteria (effectiveness, targeting, and heterogeneity) and analyze a statistical condition that is necessary but insufficient for proving expertise in interventions. We highlight a fundamental impossibility result: expertly targeted action cannot be identified from observational data alone.

\paragraph{Contribution 2: Quantitative and Qualitative Evidence of Expertise.} We provide empirical evidence for advisor expertise using both quantitative and qualitative data from the MAAPS trial. Quantitatively, we test the statistical condition characterized in our theoretical framework, whether logged advisor interventions depend on student outcomes (persistence, GPA) conditional on algorithmically available features; our analysis is relative to any potential algorithm based on these features. We identify five such potential expert interventions ($p<0.05$) (\S\ref{sec:expertise_quant}). Qualitative analysis of advisor comments shows that advisors incorporate non-algorithmic context, such as the student's personal circumstances, career goals, or financial troubles, into these interventions. Finally, we document heterogeneity in advising styles, finding that one style elicits more holistic information about students and is associated with improved graduation rates. %
Together, these findings emphasize the importance of advisor discretion for administering targeted student support (\S\ref{sec:expertise_qual}).
\vspace{2mm}
\noindent

Our paper is organized as follows: After a discussion of related work (\S\ref{sec:related_work}) and our dataset and methodology (\S\ref{sec:data_approach}),
\S\ref{sec:theory} introduces a theoretical framework for \emph{expertly targeted actions} and discusses the implications of testing for expertise using observational data. \S\ref{sec:expertise_quant} applies a statistical test of expertise to quantitative data from the MAAPS trial and identifies actions that were plausibly ``expertly targeted.'' \S\ref{sec:expertise_qual} analyzes the qualitative data to further illustrate evidence for expertise. \S\ref{sec:discussion} concludes with takeaways for the design, evaluation, and implementation of algorithmic systems. 

We recommend readers interested in the mathematical framework include \S\ref{sec:theory}. For readers primarily interested in the empirical results, \S\ref{sec:theory} can be skipped.

\section{Related Work}\label{sec:related_work}

In recent years, there has been a growing interest in the downstream effects of algorithm-assisted human decisions and the importance of human-algorithm collaboration to shaping these impacts. We review key contributions in these areas, contextualizing how they inform our study of algorithm-assisted advising.

\subsection{Causal Evaluation of Algorithm-Assisted Decision Making}

\paragraph{Predictive vs. Causal Evaluation of Algorithmic Interventions.} Predictive algorithms, such as the student risk assessment tools examined in this study, are typically evaluated by assessing their accuracy against a hold-out dataset. However, a number of factors outside of predictive accuracy can affect the ultimate impact a predictive algorithm has on outcomes: for instance, how well predictions can indicate appropriate interventions \cite{liuActionabilityOutcomePrediction2023, fernandez-loriaCausalDecisionMaking2022, barabasInterventionsPredictionsReframing2018}. Informed by these limitations of evaluating predictions alone, a growing body of literature instead evaluates algorithms through quantifying the downstream effects of algorithm-assisted human decisions using techniques from causal inference. Notably, Imai et al. \cite{imaiExperimentalEvaluationAlgorithmassisted2023} conduct the first randomized field experiment to evaluate how an algorithmic decision aid in pretrial risk assessment affects human judgment and decision quality in the U.S. criminal justice system. Follow-up work by Ben-Michael et al. \cite{ben2024does} focuses on whether algorithm-assisted human decisions improved decision-making ``accuracy'' over counterfactual baselines. Raji and Liu \cite{rajiEvaluatingPredictionbasedInterventions2025} discuss methodological challenges of study design given potential cognitive biases of a human judge.

Compared to this prior work, our study shifts the methodological focus from estimating effects of algorithmic interventions per se to the mechanisms of human discretion: we investigate whether, how, and when human decision makers contribute additional, outcome-relevant expertise that enables them to deliver appropriate interventions, improving causal impact. 

\paragraph{Causal Evaluation of Learning Analytics Interventions.} In the educational domain, the importance of studying downstream effects of algorithmic student risk assessment is well-recognized~\cite{balfanzEarlyWarningIndicators2019}. Yet, to the best of our knowledge, there has been little causal evaluation of student risk assessment. For instance, a 2019 review finds that there had been just 11 evaluations of the relationship between any learning analytics interventions and downstream outcomes in higher education, only five of which used randomized designs \cite{larrabee_sonderlund_efficacy_2019}. Three of these randomized studies examine the effect of using student risk prediction to target interventions like instructor or advisor contact with students, all finding positive impacts on academic performance \cite{jayaprakash_early_2014, lu_applying_2017, milliron_mark_david_insight_2014}. There have also been a handful of additional studies aiming to establish causal impact of student risk assessment in US primary and secondary education \cite{Faria2018GettingSO, maciverEfficacyStudyNinthGrade2019}. In a notable recent example, Perdomo et al. \cite{perdomoDifficultLessonsSocial2023} use a regression discontinuity design to estimate the causal effect of intervention targeting using student success risk scores in Wisconsin public schools, finding small effects that they suggest may be equally attainable through school-level interventions.

In general, obstacles to causal evaluation include the logistical difficulties of implementing randomization in educational institutions \cite{dawson2017from}, and collecting longitudinal data on student features, interventions, and outcomes together \cite{balfanzEarlyWarningIndicators2019}. Even when randomized or quasi-experimental designs are implemented, we often do not understand why a positive effect occurs due to a lack of qualitative data \cite{larrabee_sonderlund_efficacy_2019}. The MAAPS study is a rare exception on both fronts: a randomized trial with student features available to a predictive model and detailed structured and qualitative data on advisor decision making, over a period of four years. By studying advisor decision making in the treatment arm of an effective student risk assessment intervention, our work aims to understand the role of human expertise in driving positive impact.

\subsection{Human Discretion in Algorithm-Assisted Decision Making}

\paragraph{Information Asymmetries}  Even in settings where algorithms outperform human decision makers, individuals may possess private information that could refine
algorithmic recommendations \cite{meehlClinicalStatisticalPrediction2013}. Our approach to modeling these \textit{information asymmetries} is most similar to Alur et al.'s \cite{alurAuditingHumanExpertise2023}. Alur et al. devise a statistical test to audit for human use of ``algorithmically-unavailable information'' in improving machine predictions above any possible predictor based on a limited dataset. We apply this test to the MAAPS dataset with slight modifications. Overall, our work differs from Alur et al.’s with four key contributions. First, we generalize Alur et al.’s findings to settings where human experts use non-algorithmic information to \emph{take actions} rather than make predictions. Second, we formalize the expert decision problem using structural causal modeling and prove identifiability results. Third, we develop a novel mixed-methods analytical framework to complement the quantitative test, coding advisor comments to identify the particular non-algorithmic information guiding advisor decision making. Fourth and finally, while Alur et al. focus on medical prediction, we apply our methods to a distinct domain, educational interventions in the MAAPS study. In this setting, algorithmic context may play a unique “translation” role in bridging from predictions to interventions  \cite{liu2024bridging}: Advisors must choose among many possible interventions in order to support students, and non-algorithmic information may help them navigate this expanded action space.

In addition to Alur et al. \cite{alurAuditingHumanExpertise2023}, other work has proposed for studying complementary performance of algorithms and humans in prediction or decision tasks. For example, recent work by Hullman et al. \cite{hullman2024decision} develops a decision-theoretic framework to formalize when human decision makers can be expected to make utility-maximizing choices in algorithm-assisted contexts, which can be applied to quantify the value of information provided by such systems in synthetic decision-making tasks \citep{guo2025value}. Additionally, Rastogi et al. \cite{rastogiTaxonomyHumanML2023} develop a general framework in which to express many kinds of machine-human complementarity, including information asymmetries. Donahue et al. \cite{donahue_human-algorithm_2022} also devise a general framework for studying complementary performance. McLaughlin and Spiess \cite{mclaughlinDesigningAlgorithmicRecommendations2024} devise a causal framework for designing a recommendation algorithm in light of human patterns of compliance to achieve complementary performance. Our study differs from this prior work in its particular focus: Our framework is designed to empirically assess evidence for \textit{expertly targeted interventions} (rather then predictions, as is the main focus of \cite{rastogiTaxonomyHumanML2023} and \cite{mclaughlinDesigningAlgorithmicRecommendations2024}),  taken by a non-ideal agent in a real-world deployment of an algorithmic aid (rather than a rational, utility-maximizing agent, as with \cite{hullman2024decision}), which improve outcomes through information asymmetries (rather than more general kinds of complementarity, as with \cite{donahue_human-algorithm_2022}).

\paragraph{Reliance}

A growing body of work studies how humans combine their own judgment with the algorithmic predictions in practice, by examining what scenarios lead to humans relying more or less on algorithms and whether this improves outcomes. For example, Snyder et al. \cite{snyder2024algorithm} find that humans rely on algorithms more when the algorithms perform better than humans (on average) and when the humans are required to make decisions quickly.  Some work also explicitly examines the role of information asymmetries in reliance. For instance, Balakrishnan et al. \cite{balakrishnanHumanAlgorithmCollaboration2025} show that people may discount prediction errors by over-relying on algorithms when they have access to algorithmically-invisible context that could improve recommendations, Agarwal et al. \cite{agarwalCombiningHumanExpertise2023} find that human decision makers make errors by treating AI predictions as statistically independent from their own information, and Hoong et al. \cite{hoongImprovingAIAssistedDecisionMaking2025} find that human decision makers combine their private information with algorithmic predictions more effectively when the predictions are coarsened to lower cognitive burden. Poulidis et al. \cite{poulidisActionVsAttention2025} find evidence that prompting human decision makers with \textit{algorithmic attention signals} highlighting critical decisions rather than \textit{algorithmically-suggested actions} may improve decision-making more in the long run, since action suggestions steer decision makers into ``uncharted waters'' \cite{poulidisActionVsAttention2025}. In contrast to these studies, our goal is not to evaluate how well human agents combine AI predictions and their prior information, but instead to audit for instances of decision makers relying on non-algorithmic information. 

\subsection{Bias, Expertise and Discretion}

While predictive models are often used with the aim of closing equity gaps, their use can pose a risk of exacerbating racial biases in education. Notably, an investigation by The Markup found that many algorithmic predictors used by many universities incorporate race variables as predictors of student success, with effects like exacerbating historical biases, stigmatizing students, and raising risks like steering minority students into ``easier’’ majors \cite{feathers2021racepredictor}. A host of studies demonstrate that predictive algorithms for student success sometimes make differentially accurate predictions for disadvantaged groups \cite{penn_center_for_learning_analytics_at-riskdropoutstopoutgraduation_2025}. Additionally, scholars have analyzed other ways that algorithmic interventions in education could exacerbate inequalities beyond disparities in predictive accuracy. For instance, Baker and Hawn \cite{bakerAlgorithmicBiasEducation2022} review how bias can arise at many stages of the machine learning pipeline, from data collection to deployment. Holstein and Doroudi \cite{holsteinEquityArtificialIntelligence2021} conceptualize different ways that educational AI could exacerbate inequity by applying different ``lenses'' to analyze the socio-technical systems in which they are embedded.

Discretion can amplify or reduce bias. For example, \citet{greenDisparateInteractionsAlgorithmintheLoop2019} examine how algorithmic risk scores influence human decision making in a lab study, finding that human evaluators exhibit selective adherence to algorithmic recommendations, often amplifying existing biases. \citet{danielle2017expertise} further investigates how expertise in peer review influences evaluators' reliance on algorithmic recommendations, showing that domain experts may either mitigate or exacerbate algorithmic biases depending on their pre-existing heuristics and decision-making frameworks. \citet{mclaughlinFairnessMachineAssistedHuman2022} show that whether the algorithm amplifies or reduces the bias of the human decision-maker, with respect to a protected class, depends on whether the algorithm uses the protected class as a feature. We consider the equity impacts of discretionary uptake of algorithmic recommendations  in education  and beyond in \S\ref{sec:discussion}.

\section{Overview of Data and Methodology}\label{sec:data_approach}

This section provides an overview of the data and methodology used in our study. In \S\ref{sec:data_methods}, we describe the MAAPS randomized control trial at Georgia State University, and the structured and unstructured data sources we analyze. \S\ref{sec:mixed_methods} outlines our mixed-methods analytical framework, explaining how we integrate quantitative and qualitative approaches to study advisor discretion and expertise.
\subsection{MAAPS Experiment and Data}
\label{sec:data_methods}

\subsubsection{Experiment}
\label{sec:data_methods:experiment}

Our data comes from Georgia State University's participation in a randomized control trial run by the National Institute for Student Success (NISS) at eleven institutions from 2016 to 2022. The trial tested for effects of an intensive advising protocol called the Monitoring Advising Analytics to Promote Success (MAAPS) intervention, which has three pillars: (1) a data-driven early warning system; (2) regular review of degree plans; and (3) administering targeted and timely interventions based on regular review and early alerts.

The MAAPS data driven early alert system (1) at GSU comprised of an academic risk predictor (``predicted supported level'') that was embedded within a third party software, which advisors could access but  students could not.\footnote{EAB's \emph{Navigate} student success management platform software.} Historical risk prediction data and model details for the MAAPS cohort were unavailable for this study.

At GSU, first-year students are assigned advisors from a centralized office who provide them with academic guidance from enrollment to graduation. The MAAPS intervention was implemented among the entering class of 2016, assigning students in the treatment group to a select group of five specially trained advisors to implement the MAAPS protocol, while those in the control group were assigned to standard academic advisors.\footnote{At any one time, there were three active MAAPS advisors. Due to advisor turnover, there were five total MAAPS advisors over the course of the experiment. In fall 2017, one of the MAAPS advisors (Advisor A) left GSU and was replaced by a successor who took over until summer 2018, after which this advisor left and was replaced by another.}

At the four-year mark, students who received MAAPS advising demonstrated a  statistically significant increase in GPA (by $0.16$ points on average) and credit accumulation (by 6 credits on average) \cite{rossmanMAAPSAdvisingExperiment2021}. At the six-year mark, students in the treatment group showed a statistically significant increase in graduation rate (by $7\%$) \cite{rossmanMAAPSAdvisingExperiment2023}.\footnote{GSU was the only institution to show significant positive treatment effects of the MAAPS intervention; we discuss potential reasons for this in \S\ref{sec:discussion:implementation}.}

\subsubsection{Data}
\label{sec:data_methods:data}
We analyze data from the first four years of GSU's participation in the study in order to understand advisor's discretionary decision making in the treatment arm of the experiment. 

\paragraph{Structured Student Baseline Features and Outcomes Data.} For all 1037 students (520 in treatment and 517 in control) who participated in the MAAPS study, the dataset includes their baseline characteristics at the time of enrollment to GSU (such as their demographics, high school GPA and their scores on college entry tests, with a full list provided in Appendix \ref{appendix:dataset:baseline}), as well as timeseries data on their educational outcomes updated each term (such as their GPA and credit accumulation, with a full list in Appendix \ref{appendix:dataset:timeseries}), including their graduation outcomes. This data allows us to understand how discretionary decision making connects to students' educational outcomes and to look for heterogeneity of treatment effects in their baseline features. 

\paragraph{Structured Data on Advisor Interventions.} We also rely on logs that advisors in the treatment arm kept on all their interactions with students, providing a detailed picture of their decision-making processes. The five advisors working with treatment students in the MAAPS experiment held 5,269 meetings with students. For each meeting, advisors recorded structured data on interventions they performed during the interaction (selected from a list of 20, in Appendix \ref{appendix:dataset:meeting_logs}, for instance advising students to correct their schedule, see their instructor, or seek career counseling). We use this data to audit quantitatively for actions taken by advisors that cannot be ``explained'' by algorithmic prediction. \label{sec:data:interventions}

\paragraph{Unstructured Advisor Meeting Comments.}The advisor logs also include 1,888 free-form comments, in which advisors optionally elaborate on meetings. We use these comments to perform a complementary qualitative analysis of advisors' decision making.

\subsubsection{Data Cleaning}
\label{sec:data_methods:cleaning}

During the data pre-processing and cleaning phase, we remove students with missing data for transfer status and expected family contribution (EFC) for all quantitative analyses.  Following exclusion of students missing data on transfer status, all students are non-transfers. Following these exclusions, we have $427$ students in treatment and $423$ in control. 

For the $\text{ExpertTest}^{*}$ (\S\ref{sec:expertise_quant}), we perform additional test-specific exclusions. First, we exclude meeting data from summer semesters due to lower student enrollment and different student behaviors over summer semesters as compared to the academic year. Second, we  exclude students missing baseline attribute or outcome data. This results in a test population of $424$ students.

\subsection{Our Mixed-Methods Approach}
\label{sec:mixed_methods}
This study follows a mixed-methods approach, integrating a theoretical framework of advisor decision making (\S\ref{sec:theory}) with both quantitative (\S\ref{sec:expertise_quant}) and qualitative data analysis (\S\ref{sec:expertise_qual}). As aforementioned, primary analysis of the MAAPS RCT \citep{rossmanMAAPSAdvisingExperiment2021,rossmanMAAPSAdvisingExperiment2023}  focused on estimating average treatment effects at the population level and was purely quantitative.\footnote{While detailed data on advisor meetings was collected as part of the experiment, it was not previously analyzed. This study directly addresses this gap while introducing new research questions.}

Our mixed-methods analytical framework is designed to investigate the role of advisor discretion and expertise within the treatment group. Our motivation for integrating quantitative and qualitative methods is two-fold: (1) to develop a comprehensive understanding of advisor expertise in interventional decision making, where qualitative insights help interpret and contextualize quantitative findings; %
and (2) %
to produce novel qualitative insights that can directly inform professional development for advisors, reveal gaps, and support policy recommendations.

Our domain-agnostic approach, combining causal inference, statistical testing, and qualitative analysis, applies to studying human discretion and expertise across domains, from college advising to healthcare, social services, hiring, and other settings where understanding the interaction between human judgment and algorithmic support is critical.

We reflect below on our roles as researchers and the context in which we conducted this study.  %
Further details on our methods are in the relevant sections of the paper.

\paragraph{Researcher positionality.} As researchers, our identities, disciplinary training, and institutional affiliations shape how we approached the research questions, interpreted the data, and interacted with the broader context of advising and algorithmic decision making. The core research team (Authors 1, 4, and 5) included a graduate student and two tenure-track faculty at R1 institutions, with training in computer science, operations research, and the societal impacts of algorithmic decision making in educational applications in various U.S.-based contexts. %
The research process was shaped by a collaboration between the core research team and staff at the National Institute for Student Success (NISS) at GSU. 
From November 2023 to November 2024, Authors 4 and 5 held biweekly meetings with staff members at NISS, including advising Authors 2 and 3, to understand the MAAPS intervention and its implementation context, as well as to determine the structure and limitations of the advising data. These meetings grounded the quantitative and qualitative analyses in the realities of the program.

Access to the MAAPS data was limited to de-identified records shared under a data use agreement with NISS, and the core research team did not interact directly with students or advisors. The Princeton University and Columbia University Institutional Review Boards determined the research was exempt from further review, as it does not qualify as human subjects research. The interpretation of advising practices was shaped by the core team's outsider perspective despite best efforts to understand the unique institutional context at GSU, and by the structure and limitations of the available data. In the qualitative analysis, the team sought to preserve the nuance of advisors’ written reflections while acknowledging the constraints of interpreting them at a distance. 

Finally, as the current article constitutes a secondary analysis of the MAAPS experiment, the original design and data availability have also shaped current research questions and methods.

\section{Theoretical Framework: Expert Interventions}
\label{sec:theory}

Consider the following intuitive example of%
 algorithmically-``invisible'' information guiding discretionary action in %
advising: A student meets with their advisor, explaining that a medical condition prevents them from walking long distances. In response, the advisor works with the student to design a schedule with nearby classrooms that also meets their academic interests. The student is able to achieve a high semester GPA.

Based on a real case in the MAAPS study, this example illustrates how advisors can draw on rich, %
``non-algorithmic'' \textit{context} to target their support---in this case, a health condition that impacts the student's schedule. %
Crucially, human expertise in the interventional setting involves two components: (1) accessing this additional context and (2) also appropriately \textit{acting} on it through tailored guidance to positively influence student outcomes. This raises the question: what role does this uniquely human expertise---using contextual knowledge to guide appropriate action---play in improving outcomes?

In this section, we propose a causal framework to answer that question. We present a stylized model of human actions in the presence of algorithmic decision aids, and resulting outcomes. %
The model is inspired by and generalizes Alur et al.'s  \cite{alurAuditingHumanExpertise2023} framework for detecting human expertise in the purely predictive setting. 
While their framework address the first requirement of (1) accessing the additional context to better predict the student's state, we extend this to include the second requirement of (2) acting appropriately on the context to improve outcomes through interventions.\footnote{Alur et al. \cite{alurAuditingHumanExpertise2023} make the assumption that predictions are not ``performative'' \cite{perdomo_performative_2020}, that is, that they do not impact outcomes.}
We begin in \S\ref{sec:theory:model_and_definition} by introducing our model for algorithm-assisted interventions.~\S\ref{sec:theory:targeted_action} defines \textit{human expertise in action targeting} as the use of non-algorithmic context to select more effective actions.  In \S\ref{sec:theory:conditional_condition}, presents theoretical results on detecting the presence of expertly targeted actions from data. Finally, \S\ref{sec:theory:implications} discusses how this theoretical framework informs our quantitative and qualitative investigation of the MAAPS data.

\subsection{Model of Decision Making}
\label{sec:theory:model_and_definition}

We %
model %
algorithm-assisted decision-making processes using structural causal models (SCMs), which define the way that variables causally influence one another through a set of functional relationships \cite{pearl_causal_1995,pearl_causality_2021,peters_elements_2017,bareinboimPearlsHierarchyFoundations2022}. 

Consider a scenario where a human decision maker, at each time step $t$, needs to make a decision to try to improve outcomes at time $t+1$. We define an SCM $\mathcal{A}=\langle \mathbf{V}_\mathcal{A},\mathbf{\N}_\mathcal{A},\mathcal{F}_\mathcal{A},\mathbb{P}_\mathcal{A}\rangle$ with endogenous variables $\mathbf{V}_\mathcal{A}$, exogenous noise $\mathbf{\N}_\mathcal{A},$ structural equations $\mathcal{F}_\mathcal{A}$ and probability distribution $\mathbb{P}_\mathcal{A}$ %
to represent this single-stage decision scenario. $\mathbf{V}_\mathcal{A}$ includes $A_t$, the action taken by the advisor, and $Y_{t+1}$, the outcome at the next time step, like the student's term GPA, and may also include $\mathbf{X}_t$, any outcome-relevant administrative data about students observed by advisors (through an algorithmic decision aid),  and $\mathbf{U}_t$, any outcome-relevant non-algorithmic context at time $t$ (for instance, the information learned through advising conversations).\footnote{If there is no algorithmically-available data observed and incorporated by advisors, then $\mathbf{X}_t$ may be excluded from $\mathbf{V}_\mathcal{A}$. If there is no non-algorithmic context observed and incorporated by advisors, then $\mathbf{U}_t$ may be excluded from $\mathbf{V}_\mathcal{A}$. In these cases, Assumption \ref{assumption:dag} is adjusted to omit assumptions about edges between excluded variables and any other model variables.}

Here, outcome relevance means that a variable is an input of the process generating the student's outcomes $Y_{t+1}$. These processes are defined by functions \( f_V \in \mathcal{F}_\mathcal{A} \) for each variable \( V \in \mathbf{V}_\mathcal{A} \),  modeling \( V \) as the causal output of a subset of other variables in \( \mathbf{V}_\mathcal{A} \), and exogenous noise \( \N_V \in \mathbf{\N}_\mathcal{A} \). 

\begin{figure}[ht]
\centering
\includegraphics[width=0.6\columnwidth]{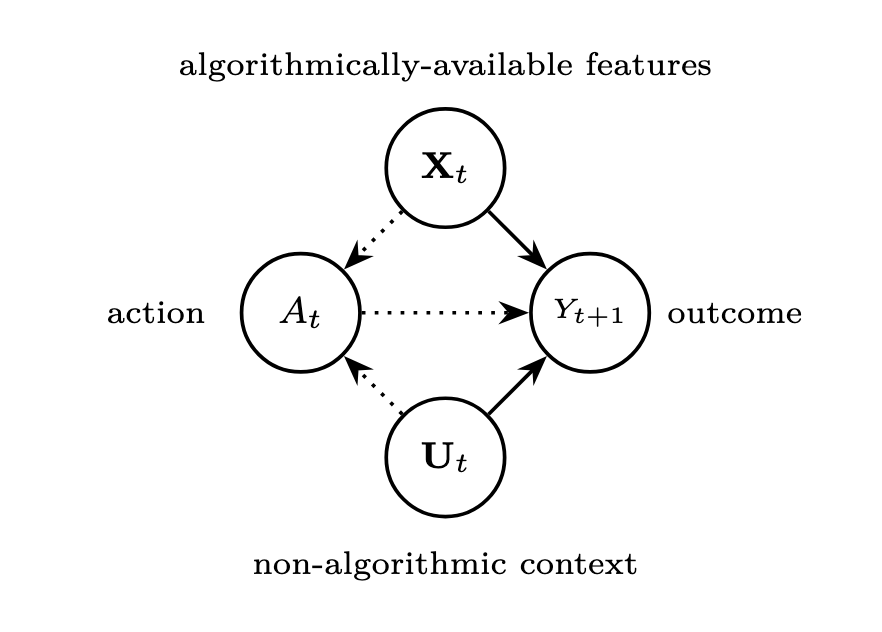}

\caption{Causal graphical model of a single-stage, algorithm-assisted interventional decision setting. Dotted edges represent potential functional dependencies.}\label{fig:graph_decisonmaking}
\end{figure}

The causal model $\mathcal{A}$ induces a directed acyclic graph (DAG) $\mathcal{G}_\mathcal{A}$ representing these processes, with nodes corresponding to endogenous model variables $\mathbf{V}_\mathcal{A}$, and a directed edge from variable \( V_1 \) to \( V_2 \) whenever \( f_{V_2} \) takes \( V_1 \) as an input. We make an assumption about the functional relationships $\mathcal{F}_\mathcal{A}$ by imposing constraints on $\mathcal{A}$'s induced graph:

\begin{assumption}[Graphical constraints]\label{assumption:dag} As depicted in Figure~\ref{fig:graph_decisonmaking}, $\mathcal{G}_\mathcal{A}$ includes directed edges from $\mathbf{X}_t$ and $\mathbf{U}_t$ to $Y_{t+1}$, may include directed edges from $\mathbf{X}_t$ and $\mathbf{U}_t$ to $A_t$, may include a directed edge from $A_t$ to $Y_{t+1}$, and has no other directed edges. %
\end{assumption}

This means the outcome is a process of both the algorithmically-available features ($\mathbf{X}_t \to Y_{t+1}$) and non-algorithmic context ($\mathbf{U}_t\to Y_{t+1}$). Advisors take actions that may be based on both these sources of information ($\mathbf{X}_t\to A_t$ and $\mathbf{U}_t\to A_t$). These actions may then influence the student’s future outcomes ($A_t \to Y_{t+1}$).

Finally, the model specifies a joint distribution $\mathbb{P}_\mathcal{A}$ over the exogenous noise variables $\mathbf{\N_\mathcal{A}}$. Together with the structural equations $\mathcal{F}_\mathcal{A}$, $\mathbb{P}_\mathcal{A}$ induces unique observational and causal distributions on all model variables.

\subsection{Definition of Expertly Targeted Action}
\label{sec:theory:targeted_action}

Suppose a MAAPS advisor recommends a particular course to a student. When does this action demonstrate expertise? We state three criteria that must be met for the action to count as expertly chosen. To simplify the definition, we consider a setting in which both the action $A_t$ and external information $U_t$ are binary variables, discussing generalizations to nonbinary contrasts in Appendix \ref{app:theory:nonbinary_contrasts}.

\begin{definition}[Expertly targeted action] \label{def:expertly_targeted_action}
For a  student with $\mathbf{X}_t=\mathbf{x}$, $U_t=u$, a human decision maker's action $A_t=1$ is \textbf{expertly targeted} when three conditions are met:

\begin{enumerate}

\item \textbf{Effective action.} Taking the action improves the outcome. (Requires $A_t\to Y_{t+1}$).
\begin{align*}
\mathbb{P}(Y_{t+1} = 1 \mid do(A_t=1), \mathbf{X}_t=\mathbf{x}, U_t=u) & > \\
\mathbb{P}(Y_{t+1} = 1\mid do(A_t=0), \mathbf{X}_t=\mathbf{x}, U_t=u)
\end{align*}
\item \textbf{Targeted action.} The human decisionmaker chose to take the action based on non-algorithmic information. (Requires $U_t\to A_t$).
\begin{align*}
\mathbb{P}(A_{t}=1|do(U_{t}=u),\mathbf{X}_t=\mathbf{x}) & > \\\mathbb{P}(A_{t}=1|do(U_{t}=1-u),\mathbf{X}_t=\mathbf{x})
\end{align*}

\item 
\textbf{Heterogeneous effects.} The relative effectiveness of the action is driven by heterogeneity in the non-algorithmic information.(Requires both $A_t\to Y_{t+1}$ and $U_t\to A_t$).

With \( \Delta(u)=\mathbb{P}(Y_{t+1}=1|do(U_{t}=u,A_{t}=1),\mathbf{X}_t=\mathbf{x}) - 
\mathbb{P}(Y_{t+1}=1|do(U_{t}=u,A_{t}=0),\mathbf{X}_t=\mathbf{x}) \),
\begin{align*}
\Delta(u)  & >  \Delta(1-u) 
\end{align*}
\end{enumerate}
\end{definition}

To count as expertly targeted, then, a course  recommendation should improve the student's GPA (effective action). The advisor should choose to recommend the course based on context unavailable to an algorithm, say the student's interests and health needs as learned in an advising conversation (targeted action). And the course should improve the student's GPA \textit{because} it meets the student's interests, not because, say, all students in the course get As (heterogeneous effects).\footnote{This definition of expertise reflects a notion of targeting as personalizing interventions based on causal heterogeneity \cite{athey_machine_2024,fernandez-loriaCausalDecisionMaking2022,shchetkina2024heterogeneity}. Similar to a definition of ``actionable heterogeneity'' proposed by Shchetkina and Berman \cite{shchetkina2024heterogeneity}, we require that heterogeneity in the targeted attribute contributes to observed improvements from the action (heterogeneous effects). However, unlike Shectkina and Berman's, our framework does not require that the ideally rational decision maker \emph{should not have} performed the action if the individual did not possess the targeted attribute; we replace this with a requirement that the actual decision maker \emph{would not have} performed the action if the individual did not have this attribute (targeted action). This choice corresponds to a difference in setting. Our goal is to identify instances of targeting in an individual agent's behavior, rather than identify whether ideal policies can benefit from targeting, as Shchetkina and Berman do.} %

\subsection{Identifying Expertly Targeted Action}
\label{sec:theory:conditional_condition}
Suppose we have an observational dataset $\mathcal{D}$ of $n$ $(\mathbf{x_i},y_i,\hat{y}_i)$, observations drawn from the joint distribution over variables $\mathbf{X}_t$, $A_t$ and $Y_{t+1}$.\footnote{This observational distribution excludes $U_t$ because it is uncommon to collect data on $U_t$, the private, algorithmically-invisible information that humans incorporate in their decision making (and moreover, it is unclear how to systematically collect  it).} Under the definitions in \S\ref{sec:theory:model_and_definition}, can we identify whether an advisor demonstrates expertise in targeting from this observational data? 

We prove an impossibility result that we cannot detect human expertise from observational data alone. However, a related statistical condition---conditional dependence of actions $A_t$ and future outcomes $Y_{t+1}$ given algorithmically available data $\mathbf{X}_t$---is a necessary condition for expertly targeted action, and sufficient for related criteria. We test whether this condition is met in the MAAPS dataset in \S\ref{sec:expertise_quant}.

\subsubsection{Assumptions}\label{sec:theory:assumptions}

In our theoretical results, we invoke three common assumptions about $\mathbb{P}_\mathcal{A}$, the distribution over the noise variables $\mathbb{\N_\mathcal{A}}$:

\begin{assumption}[Markovian condition] \label{assumption:markovity} %
The distribution $\mathbb{P}_\mathcal{A}$ is such that the noise variables $\mathbb{\N_\mathcal{A}}$ are jointly independent \citep[p.30]{pearl_causality_2021}. %
\end{assumption}

\begin{assumption}[Causal minimality]\label{assumption:minimality}
The distribution $\mathbb{P}_\mathcal{A}$ is not Markovian with respect to any proper subgraph of $\mathbf{V}_\mathcal{A}$.
\end{assumption}

\begin{assumption}[Faithfulness]\label{assumption:faithfulness}
For all disjoint subsets of variables $\mathbf{X},\mathbf{Y},\mathbf{Z}\subseteq \mathbf{V}_\mathcal{A}$, $\mathbf{X}\perp \mathbf{Y} | \mathbf{Z} \implies \mathbf{X}\perp_{\mathcal{G}_\mathcal{A}}\mathbf{Y}\ |\ \mathbf{Z}$, where $\perp_{\mathcal{G}_\mathcal{A}}$ is the relation of d-separation relative to  $\mathcal{G}_\mathcal{A}$.
\end{assumption}

The Markovian condition is a standard assumption that rules out hidden confounding between model variables \citep[p.35]{bareinboimPearlsHierarchyFoundations2022}. Assuming the SCM is Markovian ensures the SCM's DAG meets the global Markov condition for graphs, that d-separation implies conditional independence \citep[pp.101-102, p.105]{peters_elements_2017}. Though standard, the absence of hidden confounding is a relatively strong assumption.\footnote{We consider potential violations of this Markovian condition in our application of the theoretical framework to quantitative and qualitative analysis of expertise in the MAAPS experiment.} Causal minimality, by contrast, is a weak assumption, sometimes viewed as a convention, that rules out models with redundant descriptions of causal relationships \citep[pp. 107-109]{peters_elements_2017}.

Faithfulness, our strongest assumption, is the converse of the global Markov condition \citep[p.107]{peters_elements_2017} and implies causal minimality \citep[p.108]{peters_elements_2017}. Unlike causal minimality, it conveys a substantive assumption about the causal mechanisms, that causal effects do not perfectly ``cancel'' each other out so as to make causally related variables appear independent \citep[pp.107-108]{peters_elements_2017}. For instance, in our model, it could be violated if $A_t \leftarrow \mathbf{U}_t\to Y_{t+1}$ perfectly offset the effect $A_t\to Y_{t+1}$. Perfect cancellation is rare, justifying the assumption \citep[p.136]{peters_elements_2017}.\footnote{For instance, for linear models, such ``cancelling'' happens with zero probability when  coefficients are drawn randomly \citep{peters_elements_2017,spirtesCausationPredictionSearch1993}.} Since it allows for identifying graph structure from statistical \textit{independence}, it is also a common assumption for causal inference from observational data, though not as common as assuming causal mechanisms are Markovian. We note only one of our results (Proposition \ref{prop:cond_independence_necessary_condition}) depends on it.

\subsubsection{Impossibility Result on Identifying Expertise from Observational Data}
\label{sec:theory:conditional_condition:nonident}

The following proposition formalizes the nonidentifiability of expertly targeted action from observational data alone. The proof, which is presented in full in Appendix \ref{app:theory:three_scms}, presents a counterexample pair of observationally equivalent SCMs that differ in whether Definition~\ref{def:expertly_targeted_action} is met.

\begin{proposition}\label{prop:nonidentifiability} Suppose  Assumptions \ref{assumption:dag} through \ref{assumption:faithfulness} are met. \textbf{Expertly targeted action} is nonetheless impossible to detect from facts about the observational distribution over $\mathbf{X}_t$, $A_t$ and $Y_{t+1}$ alone.
\end{proposition}
\begin{proof}[Proof Sketch] 
To determine whether expertise is present, we must establish first both that $A_t$ affects $Y_{t+1}$ and that they are related by a common cause $U_t$. But in general, it is impossible to distinguish whether two variables are directly related, related by a common cause, or both using just observational data about those two variables \cite{peters_elements_2017}. We carefully construct two SCMs $\mathcal{M}_1$ and $\mathcal{M}_2$ inducing identical observational distributions $\mathbb{P}(A_t, Y_{t+1})$.\footnote{We leave $\mathbf{X}_t$ out of both models without loss of generality.}
$\mathcal{M}_1$  satisfies the definition of expertly targeted action, while $\mathcal{M}_2$ does not. %
We achieve observational equivalence by tuning the exogenous noise variables in both models to produce the same distribution over $A_t$ and $Y_{t+1}$, even though their causal structures differ.
\end{proof}

Since it is impossible to identify expertly targeted action from observational data alone, in the absence of experimentation, qualitative data on how human decision makers choose actions is crucial to understanding the deployment dynamics of algorithmic decision aids. 

\subsubsection{Necessary Statistical Condition for Expertise}
\label{sec:theory:conditional_condition:necessity}
Proposition \ref{prop:cond_independence_necessary_condition} shows that, under causal faithfulness (Assumption \ref{assumption:faithfulness}), a necessary condition for expertly targeted action is conditional dependence of the advisor's action $A_t$ and the student's outcome $Y_{t+1}|\mathbf{X}_t$.

\begin{proposition} 

\label{prop:cond_independence_necessary_condition}
Suppose the DAG induced by $\mathcal{A}$ meets the constraints in Figure \ref{fig:graph_decisonmaking}(Assumption \ref{assumption:dag}) and $\mathcal{A}$ is causally faithful (Assumption \ref{assumption:faithfulness}). Then $A_t\not\perp Y_{t+1}|\mathbf{X}_t$.

\begin{proof}
For any of the three components of the expertly targeted action definition (Definition \ref{def:expertly_targeted_action}) to be met, it must be the case that $A_t \not\perp_{\mathcal{G}_\mathcal{M}} Y_{t+1} |\ \mathbf{X}_t$ (as effective action requires $A_t\rightarrow Y_{t+1}$, targeted action requires $U_t\rightarrow A_t$, and heterogeneous effects requires both). So then by causal faithfulness, $A_t\not\perp Y_{t+1} |\ \mathbf{X}_t$.
\end{proof}
\end{proposition}

This proposition suggests a valuable method---by detecting conditional dependence, we detect a superset of actions exhibiting human expertise (putting power issues aside). 

\subsubsection{Sufficient Statistical Condition for Weakened Related Criteria} \label{sec:theory:conditional_condition:sufficient} %

While insufficient to guarantee expertise, the statistical condition can guarantee non-trivial related criteria are met. To state the result, we define criteria weakening the first two of the expertly targeted action definition, beginning with effective action:

\begin{definition} \label{def:impactful_action} An action is \textbf{impactful} if for some $\mathbf{X_t=x}$, $U_t=u$,
\begin{align*} \mathbb{P}(Y_{t+1}=1|do(A_t=1),\mathbf{X}_t=\mathbf{x},U_t=u)& \neq \\
\mathbb{P}(Y_{t+1}=1|do(A_t=0),\mathbf{X}_t=\mathbf{x},U_t=u)
\end{align*}
\end{definition}

\noindent Impactful action guarantees that actions change the outcome for some student. Effective actions, which improve the outcome, are a subset of impactful actions.

Targeted action can be similarly weakened to a criterion stating that non-algorithmic context changes the actions of the advisor for some algorithmically-indistinguishable students.

\begin{definition} \label{def:non-algorithmic_action} An action is \textbf{non-algorithmic} if for some $\mathbf{X}_t=x$,
\begin{align*}
\mathbb{P}(A_{t}=1|do(U_t=1),\mathbf{X}_t=\mathbf{x})&\neq \\
\mathbb{P}(A_{t}=1|do(U_t=0),\mathbf{X}_t=\mathbf{x})
\end{align*}

\end{definition}

\noindent 
Non-algorithmic actions are a superset of targeted actions. 

The following proposition demonstrates that meeting the conditional dependence condition $A_t\not\perp Y_{t+1} |\ \mathbf{X}_t$ provides suggestive evidence for targeting, by showing that at least one of these criteria are met. We defer its proof to Appendix \ref{app:theory:sufficiency_proof}.
\begin{proposition}
\label{prop:sufficiency}
Suppose the DAG meets the constraints in Figure \ref{fig:graph_decisonmaking}(Assumption \ref{assumption:dag}), and both Assumptions \ref{assumption:markovity} and \ref{assumption:minimality} are satisfied. If $A_t\not\perp Y_{t+1} |\ \mathbf{X}_t$, then at least one of \textbf{impactful action} (Definition \ref{def:impactful_action}) or \textbf{non-algorithmic action} (Definition \ref{def:non-algorithmic_action}) is met.
\end{proposition}

Thus, if an observational dataset meets the statistical condition, we can be certain that there are students for whom the advisor's actions are having an impact, are influenced by non-algorithmic information, or both. On the one hand, this provides another justification for using the condition to identify possible instances of expertly targeted action, without requiring a strong faithfulness assumption (Assumption~\ref{assumption:faithfulness}). On the other hand, it brings to light several key limitations in interpreting the conditional independence condition. In particular, it does not establish that the action's effect is heterogeneous with respect to $U_t$, or whether the action improves the chance of a good outcome, rather than merely changing those chances.\footnote{ 
Further, it does not establish whether both impactful action and non-algorithmic targeting are met, or just one of the two, and that if both are met, they are met for the same individuals identified by $\mathbf{X}_t$ and $U_t$.}

\subsection{Implications for Expertise Auditing}
\label{sec:theory:implications}

The theoretical framework suggests that testing for conditional dependence of advisor actions $A_t$ and outcomes $Y_{t+1}$ given algorithmically available features $\mathbf{X}_t$ allows us to identify interventions that meet a necessary condition of expertly targeted action (\Cref{prop:cond_independence_necessary_condition}) and a sufficient condition for weakened criteria (\Cref{prop:sufficiency}). %
We implement this test on the MAAPS dataset in \S\ref{sec:expertise_quant}, discussing alternative explanations for a significant result in \S\ref{sec:expertise_quant:alternative_mechanisms}.

Though quantitative observational data cannot definitively identify expertise, qualitative data can be used to provide greater insight into advisor decision making and provide evidence for expertly targeted action above alternative explanations. In \S \ref{sec:expertise_qual}, we use advisor meeting notes to look for evidence of targeted action (Definition \ref{def:expertly_targeted_action}.2), advisors using non-algorithmic context to inform their choice to take an intervention.

\section{Quantitative Findings: Empirical Evidence of Advisor Expertise}\label{sec:expertise_quant}

On paper, two students may appear indistinguishable, sharing standardized test scores, demographics, and financial aid statuses. Yet after meeting with each of them, an advisor might recommend different schedules---because the students differ in ways that are invisible from basic administrative data, such as their interests and career aspirations..

In this section, we describe results of a statistical test to identify situations like the example above in the MAAPS experiment's treatment arm. When advisors target action to students using ``non-algorithmic information'' learned in advising meetings, these targeting decisions could not be captured by \textit{any} possible algorithm based on administrative data alone. Correspondingly, we refer to actions that improve student outcomes because they were targeted based on non-algorithmic information as ``expertly targeted actions'' (formally defined in Definition  \ref{def:expertly_targeted_action}), reflecting a uniquely human contribution to advising. 

To search for evidence of expertly targeted action in the MAAPS dataset, we test for dependence of (1) advisor interventions recorded in meeting logs (denoted $A_t$) and (2) students’ future academic outcomes (denoted $Y_{t+1}) $, like their GPA, given (3) algorithmically-available features about students (denoted $
\mathbf{X}_{t}$):
\[
H_0 : A_t \perp Y_{t+1} | \mathbf{X}_t
\]

Intuitively, the condition looks at whether algorithmically-available information about students exhausts association of advisor actions and student outcomes. If not, any remaining association may reflect the advisor drawing on non-algorithmic information relevant to the outcome to target their actions to otherwise indistinguishable students, like in the example vignette.\footnote{For readers of the Theoretical Framework (\S\ref{sec:theory}), the test corresponds to the conditional dependence condition analysed in Propositions \ref{prop:cond_independence_necessary_condition} and \ref{prop:sufficiency}.} 

To implement the test, we adapt a method devised by Alur et al. \cite{alurAuditingHumanExpertise2023}, calling it the ExpertTest*. We describe the test and its application to our dataset in \S\ref{sec:expertise_quant:methodology}. Our tests find five interventions with significant results, reported in \S\ref{sec:expertise_quant:results}. However, the ExpertTest* has two key limitations. First, there are potential power limitations, as discussed in \S\ref{sec:expertise_quant:power}. Second, there are alternative explanations for conditional dependence than expertly targeted action, which we discuss in \S\ref{sec:expertise_quant:alternative_mechanisms}. Subsequent qualitative analysis in \S\ref{sec:expertise_qual} provides further support for targeting based on non-algorithmic information.

\subsection{ExpertTest*: A Conditional Independence Test for Interventions} \label{sec:expertise_quant:methodology}

To test for advisor expertise in the MAAPS experiment, we adapt the test proposed by \citet{alurAuditingHumanExpertise2023}, which audits for expertise in algorithm-assisted prediction, to our setting of algorithm-assisted \textit{interventions.} We refer to this test as the ExpertTest*.\footnote{Just as for the Theoretical Framework (\S\ref{sec:theory}), in our setting of algorithm-aided intervention, we replace the \textit{predictions} of the original ExpertTest's human forecaster with the advisor's \textit{actions} at a time $t$, $A_t$ and the predictive target $Y$ with their outcomes at time $Y_{t+1}$. We include the asterisk because of two differences in our setting: First, when predictions/actions may influence the outcome, the test is insufficient to demonstrate expertise, and second, because unlike \citet{alurAuditingHumanExpertise2023}, we also test for negative association for reasons described below.}For completeness, we briefly describe the ExpertTest* following Alur et al. \cite{alurAuditingHumanExpertise2023}, and note one point of departure from the original method.

We start by constructing a sequence of $K+1$ datasets that would be exchangeable under the null hypothesis $H_0$ given above, the first of which $
\mathcal{D}_0=\{ (x_t^1,a_t^{i,1},y_{t+1}^1) \ldots (x_t^n,a_t^{i,n},y_{t+1}^n)\}$ is the true dataset, and the other $K$ of which are synthetic datasets $\mathcal{D}_1 \ldots \mathcal{D}_K$. To construct those synthetic datasets, we identify $L$ pairs of individuals with (nearly) identical $\mathbf{X}_t$ values (``algorithmically-indistinguishable individuals'') in $\mathcal{D}_0$.\footnote{When individuals in $\mathcal{D}_0$ cannot be paired exactly, as is the case for our dataset, the test can be relaxed to accommodate approximate pairing, introducing error that scales in match quality (measured by a distance function on $\mathbf{X}_t$). We refer readers to \cite{alurAuditingHumanExpertise2023} for the justification of the relaxation as well as rigorous statistical justification of the test.} Then, using those $L$ pairs, $K$ synthetic datasets are constructed from $\mathcal{D}_0$ by making swaps of the actions advisors take for paired individuals with probability $\frac{1}{2}$.  Then, using those $L$ pairs, $K$ synthetic datasets are constructed from $\mathcal{D}_0$ by making swaps of the actions advisors take for paired individuals with probability $\frac{1}{2}$.

The mean squared error of a dataset is defined as $MSE(\mathcal{D}) = \sum_{j=1}^n (a_t^{i,j}-y_{t+1}^{j})^2$ with $n=|\mathcal{D}|$. 
Let  $1[\alpha \gtrsim \beta]=1$ if $\alpha>\beta$ with ties broken randomly. Given exchangeability, the order statistics over the mean squared error $MSE(\mathcal{D})$ of these datasets, 
\[\tau_K = \frac{1}{K} \sum_{k=1}^K 1[MSE(\mathcal{D}_0)\gtrsim MSE(\mathcal{D}_k)] \]

will be distributed uniformly over $\{0,\frac{1}{K},\frac{2}{K}\ldots\}$. If the null is true, then, $\mathbb{P}(\tau_K \leq \frac{J}{K}) \leq \frac{J+1}{K+1}$ for $J\leq K$. For $\tau_K=\frac{J}{K}$, then, we can reject the null with $p$-value $\frac{J+1}{K+1}$. Since $\frac{J+1}{K+1}\approx \tau_K$, we reject the null when $\tau_K$ is small.

Intuitively, the test assesses whether advisor actions predict outcomes in ways unexplained by algorithmically-visible features $\mathbf{X}_t$. Randomly swapping actions among individuals with identical  $\mathbf{X}_t$ should not systematically worsen predictive error---as measured by $MSE(\mathcal{D})$--- unless the original actions contained additional information about outcomes, whether through access to non-algorithmic information $U_t$, a causal effect of actions, or both.

The ExpertTest* only rejects the null whenever the mean squared error in the synthetic datasets is routinely \textit{higher} than the the mean squared error of the original dataset. In Alur et al. \cite{alurAuditingHumanExpertise2023}, this one-sided test of predictions is sensible, as it ensures the test only detects instances where non-algorithmic information is used to improve the predictions of the human forecaster. But we are also interested in negative association, as actions intended to assist at-risk students may be \emph{negatively associated} with outcomes due to selection bias for hidden risk factors (see \S\ref{sec:expertise_quant:alternative_mechanisms}). This is a key point of departure from the pure prediction setting \citep{alurAuditingHumanExpertise2023}, where good predictions must be positively associated with the outcomes. To address this, for every action $A^{i+}_t$ (e.g., recommending a schedule), ExpertTest* is also applied to the transformed action variable \( A_t^{i-} = 1 - A_t^{i+}\) (e.g., not recommending a schedule). This is equivalent to defining
\[
MSE^{-}(\mathcal{D}) = \sum_{j=1}^n \left((1 - a_t^{i,j}) - y_{t+1}^j\right)^2
\]
and running the procedure described above using \(MSE^{-}\), allowing us to detect cases where taking an action is associated with worse outcomes, consistent with effective targeting of hidden risks.\footnote{Another reasonable choice would be to run a two-tailed test on the order statistic $\tau_K$, which we opt against primarily for reasons of interpretability. See Appendix \ref{appendix:cond_indep:test_methodology} for a comparison; the methods yield similar results.}

\subsubsection{Application to MAAPS Data}
The ExpertTest* is framed in terms of individual actions at a single timestep $t$ affecting outcomes at $t+1$, so to the multistage, multi-intervention nature of collegiate advising complicates variable definition. We run different variants of the ExpertTest* to accommodate these complexities. Below we describe how we define $\mathbf{X}_t$, $A_t$, $Y_{t+1}$, and $n$ for these tests. All test parameters and variables, as well as details of how we chose test parameters $L$ and $K$, are summarized in Appendix \ref{appendix:cond_indep:test_parameters}.

\paragraph{Algorithmic Information.} For all the tests, we take $\textbf{X}_t$, the algorithmic information, as students' attributes at baseline excluding their race and ethnicity, since Georgia State University excludes this information from all its early warning systems \cite{rossmanMAAPSAdvisingExperiment2023}. Following our data cleaning (\S\ref{sec:data_methods:cleaning}), the population consists of 424 students having 3,507 meetings over four years (Appendix Figure \ref{fig:meetings_per_semester}).

The test requires matching students with approximately identical algorithmically-available features. We pair students greedily using a Euclidean distance function applied to fixed-range scaled features (Appendix Figure \ref{fig:euclidian_distance}).%
We select the number of pairs for each test, $L$, to yield an acceptable tradeoff between test power, which increases in $L$, and test error, which increases in match distance (Appendix \ref{appendix:cond_indep:test_parameters}).

\paragraph{Actions.}
We draw our action data from the lists of advisor interventions (described in \S\ref{sec:data:interventions}). 

To avoid an exponential blowup of actions, we binarize actions as any advisor applying intervention $i$ in a meeting with the student from the list of 20 over a particular timescale $t$. We denote this binary variable by $A_t^{i+}$. We pool intervention data from all advisors in order to capture potential use of external information by any advisor. Because of inconsistencies in the quality of Advisor 1's intervention data (Appendix \ref{app:data:quality}), we re-run the test excluding students assigned this advisor and confirm the results' consistency.

As discussed in \S\ref{sec:expertise_quant:methodology}, we also run each test on  $A_t^{i-}=|1-A_t^{i+}|$, which detects cases in which  taking the action predicts negative outcomes.

We test each intervention variable separately for simplicity, but in practice, interventions are often applied together by advisors. Their combined application may be important for their effectiveness or targeting. To account for this, we examine correlations between actions the ExpertTest* detects as significant, summarizing in Appendix \ref{app:rq1_quant:correlations}.%

\paragraph{Outcomes.} Of the three outcomes we examine, persistence, graduation and GPA (at various timescales), all but GPA are naturally binary; we binarize the GPA outcome by testing for GPA above the mean of 3.5, testing more thresholds for robustness.\footnote{We note that the MSE here differs from its normal interpretation---whereas MSE is normally a comparison of predictions of an outcome to that outcome, here we look at whether \textit{actions} predict binary outcomes. Future work might explore alternate test statistics or conditional independence tests from \citet{alurAuditingHumanExpertise2023}'s.} %

\paragraph{Timescale.} The timescales of our action and outcome variables are chosen to avoid the introduction of feedback loops that could arise due to the multistage nature of the MAAPS experiment.\footnote{For instance, if $A^i$ is having intervention $i$ performed at any point in 4 years, and $Y$ is persistence throughout those same 4 years, this introduces a backward edge $Y \rightarrow A^i$ because students who drop out will not meet with their advisors.} We run tests over two different timescales:

\begin{enumerate}
\item Semesterly tests: Among enrolled students in each semester (Appendix Figure \ref{fig:enrolled_per_semester}), we look at the result of the application of each intervention $i$ in a meeting that semester on  term GPA or persistence to the next term (for spring terms, persistence is defined as enrollment/graduated status in either the summer or fall). In these tests, students' semester of enrollment is added as a feature to their baseline attributes $\mathbf{X}_t$, and cross-semester matches of students are prohibited.
\item Early-intervention tests: Among students enrolled for the  first year ($n=396$), we look at the result of each intervention $i$ that year on four-year graduation and four-year persistence. 

\end{enumerate}

\paragraph{Conditional Independence Test.} We test the null \[
H_0: A_t^i \perp Y_{t+1} |\ \mathbf{X}_t
\]
for each action $i$ (with both $A_t^{i+}$ and $A_t^{i-}$), timescale $t$, and outcome $Y$ described above.

\paragraph{Adjustment.} To account for multiple testing (40 hypotheses for each timescale and outcome, corresponding to $A_t^{i+}$ and $A_t^{i-}$ for $i\in [1,20]$), we adjust $p$-values for each timescale and outcome using the Benjamini-Hochberg procedure.

\subsection{Results}
\label{sec:expertise_quant:results}

Following correction for multiple testing, four interventions demonstrate significant conditional dependence with term GPA (Table \ref{table:expert_test_semesterly_gpa}) and one intervention demonstrates significant dependence with persistence (Table \ref{table:expert_test_semesterly_persistence}). Significance levels for all 20 interventions are given in Appendix \ref{appendix:quant_expertise:results}, alongside robustness checks  with different choices of threshold for GPA, and with exclusion of Advisor 1, who has inconsistent quality of intervention logs.

These five significant interventions comprise 5,615 of the 8,385 total interventions taken by advisors following our data cleaning process. Accordingly, we estimate that 67\%, or two-thirds, of interventions taken by MAAPS advisors are potentially expertly targeted. We provide this numerical estimate of the pervasiveness of significant interventions, with the caveat that a significant result is necessary but insufficient for expertly targeted action (which we address by providing qualitative evidence of expertise in \S\ref{sec:expertise_qual}).

Two of these significance results have a positive direction, indicated by a $(+)$ in Tables \ref{table:expert_test_semesterly_gpa} and \ref{table:expert_test_semesterly_persistence}, with the $p$-value interpreted as the probability that actions taken by advisors that cannot be explained by administrative data predict a \textit{positive} outcome for the student (high GPA or next-term enrollment) just by chance. The rest have a \textit{negative} direction, notated by $(-)$, with $p$-values interpreted as the probability that actions  predict negative outcomes just by chance. Negative association can be explained by selection bias for students with non-algorithmic risk factors, who may be less likely to succeed on average (see \S\ref{sec:expertise_quant:alternative_mechanisms}).

We note that the tests of first-year interventions on a four-year timescale do not meet the significance threshold (see results in Appendix \ref{appendix:quant_expertise:results}). We suspect this is due to the long time range between cause and effect: Advisors' actions in the first year are unlikely to contain detectable signal about outcomes three years later.

\begin{table}[t]
\centering
\begin{center}
\textbf{GPA $\geq 3.5$: top five interventions}
\end{center}
\begin{tabular}{l|c|c}
Intervention & Direction & BH corrected $p$-value \\
\hline
20 (Develop Degree Map) & (+) & 0.039960*\\
16 (Schedule Appointment) & (-) & 0.039960* \\
2  (Correct Schedule) & (-) & 0.049950* \\
4 (Progression Issues) & (-) &  0.049950* \\
11 (Career Counseling) & (+) & 0.053280
\end{tabular}
\caption{\normalfont Top five most significant interventions for the ExpertTest* for  term GPA $\geq 3.5$ (the mean student GPA).}\label{table:expert_test_semesterly_gpa}
\end{table}

\begin{table}[t]
\centering
\begin{center}
\textbf{Persistence: top five interventions}
\end{center}
\begin{tabular}{l|c|c}
Intervention & Direction & BH corrected $p$-value \\
\hline
17 (Recommend Schedule) & (+) & 0.039960*\\
20 (Develop Degree Map) & (+) & 0.119880\\
16 (Schedule Appointment) & (-) & 0.333000 \\
6 (See Instructor) & (-) & 0.399600\\
1  (Seek Academic Support) & (-) & 0.399600
\end{tabular}
\caption{\normalfont Top five most significant interventions for the ExpertTest* for  persistence to the next term.}\label{table:expert_test_semesterly_persistence} 
\end{table}

\subsection{Discussion: Test Power}
\label{sec:expertise_quant:power}

Interestingly, there is a higher number of significant interventions for high term GPA than for persistence. We hypothesize two primary reasons for this trend. First, the effects of actions advisors take to promote persistence may surface over longer time ranges, whereas semester GPA may be more easily impacted by advisors' recommendations in a particular term (like recommendations to take particular courses). Second, the power of the ExpertTest* is decreased for outcomes with high or low base rates. Since the test generates synthetic datasets that ``swap'' intervention values for similar students and compares them to outcomes with the mean squared error, for rare outcomes, there are fewer swaps that can affect the mean squared error of a synthetic dataset, as most individuals' outcome values are already equal (Appendix \ref{appendix:quant_expertise:power_explanation}). 

Just like with outcomes, the ExpertTest* may not identify expertly targeted \textit{interventions} that are very rare. Accordingly, Appendix Table \ref{appendix:table:swaps} shows that rare interventions in the MAAPS dataset have few swaps between algorithmically-indistinguishable pairs that could change mean squared error, reducing power  for tests of these interventions.

\subsection{Discussion: Alternative Explanations and Link to Qualitative Analysis}\label{sec:expertise_quant:alternative_mechanisms}

As our Theoretical Framework shows (\S\ref{sec:theory:conditional_condition}), conditional dependence of actions and outcomes is only a necessary condition for expertise and the five significant interventions need not be truly expertly targeted. Conditional dependence of outcomes and actions can have alternative explanations than expertly targeted action.

This fundamental limitation of quantitative analysis underscores the value of our mixed-methods approach. While statistical tests alone cannot definitively establish expertise, qualitative analysis of the advisor meeting notes can complement our understanding of the decision-making process.

In \S\ref{sec:expertise_qual}, we use qualitative data to provide evidence for the second component of expertly targeted action, called targeted action  (Definition \ref{def:expertly_targeted_action}.2), which requires that advisors use non-algorithmic information to decide what action to take. We examine whether comments for significant interventions show evidence of targeting based on non-algorithmic context. For significant interventions that are \emph{negatively} associated with the outcome, we additionally look for evidence they are targeted based on \textbf{non-algorithmic risk factors}, which place the student at risk for worse outcomes.

Finally, to the extent possible, we look for evidence for and against three alternative explanations of significant results:
\paragraph{Alternative Explanation 1. Nuisance Confounding.} Nuisance confounding between outcomes and actions could contribute to or entirely explain  a significant result. For instance, if $A_t$ is an advisor making a schedule recommendation, and $Y_{t+1}$ is next-term enrollment, their relationship may be confounded by students requesting scheduling advice---since students who request such advice are disproportionately likely to re-enroll.\footnote{Formally, this is a violation of Assumption \ref{assumption:markovity}, the absence of hidden confounding.}
\paragraph{Alternative Explanation 2. Absent Targeting.}  Even without nuisance confounding, a significant result still does not imply the presence of targeting. The action may be untargeted, with dependence explained by the action's effect on the outcome.\footnote{Formally, this corresponds to the failure of the targeted action (Definition \ref{def:expertly_targeted_action}.1)   component of the expertly targeted action definition.}

\paragraph{Alternative Explanation 3. Ineffective Action or Lack of Heterogeneous Effects.}  To qualify as expertly targeted, an intervention must actually help students to whom it is targeted. Yet a significant intervention may be targeted based on non-algorithmic information but either harm or fail to affect the outcomes of the targeted student. Alternatively, the action may impact students positively, but not in virtue of the targeting---for example, by helping all students equally.\footnote{Formally,  these possibilities correspond to the failure of the effective action (Definition \ref{def:expertly_targeted_action}.1) and heterogeneous effects (Definition \ref{def:expertly_targeted_action}.3) components of the expertly targeted action definition, respectively.}

\section{Qualitative Analysis: Evidence for Non-Algorithmic Targeted Action}
\label{sec:expertise_qual}

We conduct a two-stage qualitative analysis to better understand advisor discretion and intervention practices, complementing our quantitative findings. The first stage involves open coding to identify themes across all advisor comments, to familiarize ourselves with the data and its limitations. The second stage focuses on a targeted analysis of comments linked to statistically significant interventions from the ExpertTest* to investigate potential evidence of the \emph{targeted action} component of expertly targeted action (\Cref{def:expertly_targeted_action}). In this section, we describe the methods for our qualitative analysis in \S\ref{sec:expertise_qual:methods} and how they complement our quantitative findings about advisor expertise in \S\ref{sec:expertise_qual:results}. In \S\ref{sec:expertise_qual:styles}, we discuss results from the exploratory qualitative analysis in connection to a heterogeneous treatment effect analysis in advisor assignment.

\subsection{Qualitative Methods}\label{sec:expertise_qual:methods}

\subsubsection{Exploratory Coding}
In the first stage of our qualitative analysis, we conduct open coding to identify themes across all advisor comments and develop familiarity with the data. Following removal of students due to our data cleaning (\S \ref{sec:data_methods:cleaning}), we code a total of 1,661 comments %
left by advisors on meetings with 422 students in the treatment arm. Using the \emph{Atlas.ti} software, Authors 1 and 5 apply codes for discussion of academic matters (such as course selection and academic performance), nonacademic matters (such as financial aid or family circumstances), and relationship-building matters (such as advisors' observations on student affect or expressions of empathy for student circumstances). 

We note that, while these codes provide  insight into advisor styles and the content of meetings, they are necessarily limited to what advisors chose to document, and may not fully capture the scope or nuance of each interaction. Details on our coding process and codebook definitions are provided in Appendix~\ref{app:qual_exploratory:methods}.

\subsubsection{Coding for Expertise}

Our second round is a targeted analysis of comments to investigate potential evidence of expert targeting. We focus this stage on comments of meetings where advisors reported taking at least one of the five statistically significant interventions (\S\ref{sec:expertise_quant:results}). Using a spreadsheet, Author 1 first determines whether advisors explicitly mention the relevant intervention and report a clear reason for taking the intervention, which qualifies the comment for inclusion in the analysis. 
For recommending a schedule (Intervention 17), we additionally code for customization, identifying whether advisors tailored the content of their recommendations based on student-specific context. These criteria are rigidly enforced to mitigate the risk of inferring non-algorithmic targeting due to confirmation bias. For comments meeting the criteria, Author 1 then codes the stated reasons motivating each intervention inductively and merges them into standardized codes. These codes of advisor-provided reasons are assessed for evidence of plausibly non-algorithmic factors informing intervention targeting. Full code definitions and examples are provided in Appendix~\ref{appendix:qual_deets:qual_interventions}.

To reduce confirmation bias in identifying human expertise,  
we adopt a conservative approach: we interpret advisor reasoning only when clearly stated and apply the same coding procedures to a sample of statistically insignificant interventions for comparison.\footnote{We select the two most insignificant interventions on the ExpertTest* for each outcome with greater than 100 instances of application; one of these overlaps for both outcomes, leading to three total interventions.} %

\subsection{Qualitative Analysis of Expertly Targeted Action}\label{sec:expertise_qual:results}

\subsubsection{Evidence of Targeted Action for All Significant Interventions} 

\begin{table}[ht]
\centering
 \resizebox{0.9\textwidth}{!}{
\begin{tabular}{r|r|c|c|c|c|c}
\hline
& & \textbf{Int 2 ($-$)} & \textbf{Int 4 ($-$)} & \textbf{Int 16 ($-$)} & \textbf{Int 17 ($+$)} & \textbf{Int 20 ($+$)} \\
& & \footnotesize{Correct Schedule} & \footnotesize{Degree Progression} & \footnotesize{Schedule Meeting} & \footnotesize{Recommend Schedule} & \footnotesize{Degree Map}\\
\hline
1& Course Credits/Placement & 3 &  4 & -  & - & 1 \\
2& Course Problem & 11 & 6  & 6 & - & 1 \\
3& Degree Progress & - & 8 & - & - & 2 \\
4&Load Management & 18 &  - & - & - & - \\
5&Major Choice & - &  8 & 18 & - & - \\
6&Personal Circumstances & - &  2 & 5 & - & - \\
7&Low Engagement & - &  - & 109 & - & - \\
8&Scholarship Process & 2 &  - & 1 & - & - \\
9&Student-Initiated & 3 &  - & 8 & $243^{(*)}$ & 1 \\
10&Transfer/Dropout & - &  3 & 3 & - & - \\
11&Unregistered & - &  - & 12 & - & - \\
\end{tabular}}
\caption{\normalfont %
\textbf{Reasons found in advisor comments associated with each of the five significant interventions of the ExpertTest*.} We indicate how frequently each reason was mentioned by advisors for applying a particular intervention. Most interventions show diverse reasons behind their application. Additionally, interventions associated with \emph{negative} outcomes, noted with a $(-)$, all show evidence of targeting based on risk factors---such as \emph{Course Problem, Degree Progress, Load Management, Low Engagement, Transfer/Dropout}, and \emph{Unregistered} (defined in Appendix \ref{appendix:qual_expertise:methods}). Note the exception of Intervention 17:  whenever an explicit reason for \emph{Recommending a Schedule} was mentioned, it was recorded as \emph{Student Initiated}; however, advisors often incorporated non-algorithmic context into the \emph{contents} of the schedule (found in 81 comments; SI \ref{app:qual_expertise:results:int_17}). %
}\label{table:reason_counts} 
\end{table}

We use the qualitative coding procedure described above to determine whether advisors target significant interventions to the student based on non-algorithmic information. We include selected excerpts from advisors' comments for illustration.\footnote{Comments are paraphrased as needed for clarity and to preserve anonymity. To protect student privacy, all gendered pronouns in comments have been changed to they/them.}

Across significant interventions, we see a pattern: \textbf{advisors incorporate non-algorithmic context into their decisions about when and how to intervene.}\footnote{As discussed further in \S\ref{sec:expertise_qual:alternative_explanations}, Intervention 20 (being advised through in-person development of the degree map/planner) is an exception. We note that the absence of comments demonstrating targeting does not necessarily imply a lack of expert targeting.} For example, in providing schedule recommendations (Intervention 17), which is positively associated with persistence ($p=0.039$), advisors consider diverse factors like student health conditions, motivation to stay in college, and plans to transfer to customize their particular schedule recommendations:

\begin{quote}
``Student has [a health] condition and has to make sure all of their courses are in proximity so as to avoid walking too much.''
\end{quote}
\begin{quote}
``Student is in college because of pressures from their father, who could not attend college. Because they are covered by the GI Bill, the student did say they would give this first year their all. So my goal is to get them into classes next semester that will inspire them.''
\end{quote}
\begin{quote}
``Student wants to transfer to GT [Georgia Tech]. Says they have conditional acceptance after 30 credit hours. I explained that not all CS courses transfer to GT and they need to find out what transfers. For spring, they want to take 18 credits to ensure they will the reach 30 credit hours.''
\end{quote}

All of these comments demonstrate examples where non-algorithmic context informs advisor actions. For example, in the third comment, the context $\mathbf{U_t}$ is the student's requirements to transfer to Georgia Tech, and the $A_t$ is designing a schedule to contain 18 hours.

In many cases, advisors incorporate context $\mathbf{U_t}$ that function as \textbf{non-algorithmic risk factors}, signaling that a student may be at risk of worse outcomes. Consider \emph{recommending the student to schedule a meeting} (Intervention 16), which is negatively associated with a high GPA ($p=0.039$). Most often, advisors take this action for students because of the context $\mathbf{U_t}$ that these students failed to schedule appointments or miss their scheduled appointments, an intuitive risk factor for poor outcomes:
\begin{quote}
``Student did not attend appointment. I called student to reschedule; student stated they are having to meet with financial aid due to their refund check sent to wrong bank.''
\end{quote}

Advisors also take this intervention for students with exceptionally difficult personal or academic circumstances:
\begin{quote}
``They are currently living with friends and family. [Redacted], and because of such, they do not have a permanent place to call home.''
\end{quote}
\begin{quote}
``Student was dropped from courses back on Jan 18th due to non-payment. Student had not been attending courses the whole term. Student explained when we met the first time that they did not have the books.''
\end{quote}

The targeting based on non-algorithmic risk factors $\mathbf{U_t}$ may explain why Intervention 16 is negatively associated with GPA. The other two significant interventions associated with a low GPA---recommending a schedule correction (Intervention 2) ($p=0.049$) and advising about progression issues (Intervention 4) $(p-0.049)$---show a similar pattern of targeting based on risk factors such as course performance, degree progress, and challenging personal circumstances, as Table \ref{table:reason_counts}, which includes systematic counts of reasons advisors take each intervention, demonstrates.

Finally, our qualitative analysis suggests it would be difficult to capture all this rich context \emph{without human expertise}. In particular, dataset augmentation and algorithmic enhancements will likely be insufficient, for two reasons. First, for any given intervention, \textbf{the distribution of kinds of information that can persuade the advisor to act has a long tail:} in most cases, there is a broad spectrum of information that can lead an advisor to take a particular action. Table \ref{table:reason_counts} demonstrates the diversity of the reasons advisors cite for taking significant interventions. When advising a schedule correction (Intervention 2), for instance, advisors consider matters including student's course preferences, performance,  financial aid requirements,  projected course or extracurricular loads, prior placement exam performance, and credits earned elsewhere (which often have not been recorded in administrative data at the time of their conversation). 

Second, \textbf{subtle nuances in context, which may be flattened or ignored by tabularization, are consequential to the actions advisors take.} Consider the following pair of examples for schedule corrections (Intervention 2):

\begin{quote} 
``Student stated that they are NOT feeling comfortable in their math course. This is a course that they need special attention on and they would like to take it in the summer.  Because they have already taken two classes in the summer, I felt comfortable advising them to withdraw from the course.''
\end{quote}
\begin{quote}
``Student needs to add more credits to fall. Offered [MATH XXX]. Student didn't want to take [MATH XXX] because they said they needed a `break' from math. However, this will help the student stay on track, so I strongly encouraged them to take the class in the fall.''%
\end{quote}
These two situations, in which a student wants to postpone a math course, appear similar at first but differ in important ways that lead to different advisor recommendations. The first student has a clear plan to make up the course and a track record of success in summer courses. The second student does not offer a clear reason for postponing the math course and needs more courses, suggesting they would not benefit in the same way from a postponement recommendation. Adequate dataset augmentation may require taking such complex signals of student motivation into account.

\subsubsection{Alternative Explanations: Evidence of Confounding}\label{sec:expertise_qual:alternative_explanations} %
Conditional dependence is only a necessary condition for expertise. As summarized in \S\ref{sec:expertise_quant:alternative_mechanisms}, there are three alternative explanations for significant results.%
We ask whether our qualitative data support %
any of these alternative explanations, and  find evidence of nuisance confounding, especially in \emph{recommend schedule} (Intervention 17).

\paragraph{Some Interventions are Confounded by Nuisance Factors.} While significant interventions show signs of non-algorithmic targeting, \textbf{some additionally show signs of a confounding effect, in which advisors perform interventions after students explicitly request they do.}\footnote{We note the presence of confounding does not rule out the simultaneous presence of expertise.}

The clearest such case is recommending a schedule (Intervention 17), which positively predicts next-term persistence ($p$=0.039). Consider the following examples:
\begin{quote}
``NEUR student wanted to talk about spring semester and what to expect.''
\\\\
``Student came in the office to discuss schedule for SU '17 and FA'17 semester.''
\end{quote}
Such student requests, of which there are 243 examples, confound the relationship between actions and outcomes: A student requesting schedule recommendations for next semester from their advisor likely also intends to enroll regardless of the advisor's particular recommendations. At a much smaller scale, similar confounding also occurs in comments related to schedule corrections (Intervention 2), progression issues (Intervention 4), and being advised to schedule an appointment (Intervention 16), as the counts in row 9 of Table \ref{table:reason_counts} (student-initiated interventions) demonstrate.

\paragraph{Evidence Against Absent Targeting for Most Interventions.}
For meeting comments associated with four of the five significant interventions, we see  comments where advisors incorporate non-algorithmic information into their recommendations. This provides evidence against the absence of targeting altogether. For the remaining significant intervention, \emph{advise through in-person development of the degree map or planner} (Intervention 20), we do not have enough data to determine whether the action is non-algorithmically targeted. Advisors only explicitly describe their reason for developing the degree planner in 6 out of 284 comments associated with meetings in which the intervention was performed (Appendix Table \ref{app:table:qual_interventions_summary}).
\vspace{-1mm}
\paragraph{Inconclusive Evidence about Ineffective Action or Lack of Heterogeneous Effects.}
 The qualitative analysis does not provide adequate evidence about \emph{ineffective action} or \emph{lack of heterogeneous effects} as an alternative explanation. Comments cannot reveal how advisor interventions and non-algorithmic information causally affect outcomes; they can only tell us \emph{how advisors choose actions} to take. We note that many of the above instances of targeting that we identified plausibly are helpful to students and use non-algorithmic context relevant to the effect of the action. For example, take the case of the student with a health condition whose schedule must be adjusted so their classes are in close proximity. It is plausible that this schedule customization  helped the student, and that the incorporation of context about their medical condition was important to the intervention's success.

\subsection{Qualitative Findings on Advisor Styles}\label{sec:expertise_qual:styles}

Our exploratory qualitative analysis of advisors' comments reveals notable differences in their advising styles. Out of three advisors with a substantial numbers of comments available for analysis, while Advisor 1 and 2's comments show what we term an \emph{academic} style (Appendix \ref{app:qual_exploratory:results}) focusing on topics like schedule planning and course performance, Advisor 3's comment reflect what we call a \textit{holistic style} of broader engagement with the student%
, addressing not only academic planning, but also discussion of non-academic factors like career goals and family circumstances, observations about student personality, and expressions of empathy for student circumstances.

\begin{figure}[ht]
    \centering
    \includegraphics[width=0.5\columnwidth]{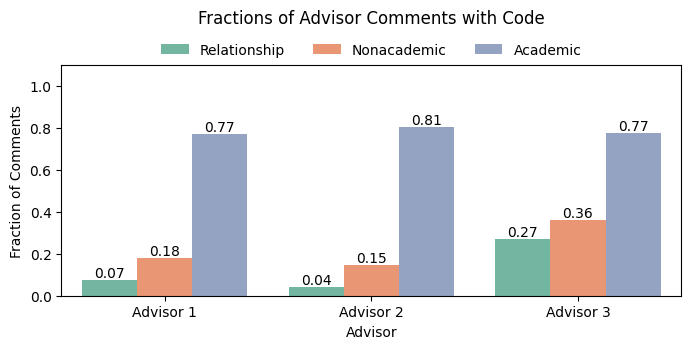}
    \caption{Fraction of meeting comments coded with academic, non-academic, and relationship codes (Appendix \ref{app:qual_exploratory:methods}), broken down by advisor. Advisor 3's comments more often bring up non-academic topics and show more signs of building an advising relationship with students. Note that fractions for different code categories within each advisor may not sum to 1 as multiple codes may be applied to one meeting.} 
    \label{fig:code_distrib}
\end{figure}

This difference is visualized in the distribution of codes applied to their meetings  (Figure \ref{fig:code_distrib}), which shows that Advisor 3's comments display a more even balance of academic, non-academic (discussions of topics like career aspirations, finances, or family circumstances), and relationship-building (expressions of empathy or observations about students' personalities) codes. A full description of the codes is included in Appendix \ref{app:qual_exploratory:methods}.

\paragraph{Association with Graduation Outcomes.} To understand whether this difference in advising style is consequential, we use MAAPS trial data to look for heterogeneous treatment effects of the MAAPS intervention across advisors on \emph{four-year graduation} (Appendix \ref{app:heterogeneity:methods}). While we do not find statistically significant differences in effect size, logistic and linear regression models produce a consistent ranking of advisor-specific coefficients with Advisor 3 at the top (Appendix \ref{app:heterogeneity:results}). This ordering provides preliminary quantitative evidence that the \textbf{holistic advising style may be more effective than the academic one.}

We hypothesize that this heterogeneity may be partly explained by the \textbf{more holistic advising style enabling Advisor 3 to expertly target student support better than the academic style.} For example, one student confides in Advisor 3 about their mental health:
\begin{quote}
``Student is very anxious about GSU, college and career in general ... They are worried they won't be successful and that they won't find a job and that they'll want to change their major.  I asked them to look up [people in their profession] and see what they did to get their jobs. The student is worried they won't get exposure to their field, so I challenged them to get exposure. They did ask about the counseling center.''
\end{quote}%
The emotional connection allows Advisor 3 to learn important non-algorithmic information $\mathbf{U}_t$ (student anxiety) to take action $A_t$ (in this case, helping the student research their chosen career path and recommending mental health resources) that is plausibly important to improving the student's outcomes.

In another example, Advisor 3 plumbs a student's deeper motivations for pursuing a nursing degree:
\begin{quote}
``Student is in the pre-nursing program, but they stated that their math and chemistry prerequisites were giving them trouble. I explained the GPA requirements. They want to help people, but I challenged them to think of why and how.''
\end{quote}
The comment shows Advisor 3 incorporating knowledge of the student's underlying altruistic motivations into their recommendation to explore other, more academically feasible paths that fulfill this aim.

\begin{table}[ht]
\centering
\begin{tabular}{c|c|c}
\hline
\textbf{Advisor} & \textbf{\# Comments} & \textbf{Average JSI} \\ \hline
Advisor 1 & 770    & 0.5960                                  \\
Advisor 2 & 364    & 0.7694                                      \\
Advisor 3  & 360      & 0.4688                   
\end{tabular}
\vspace{1mm}
\caption{\normalfont Average pairwise Jaccard Similarity Index (JSI) on qualitative codes applied in the meetings for each advisor.}
\label{tab:jaccard_codes}
\end{table}

Finally, echoing the finding that there is a ``long tail'' of diverse non-algorithmic context that advisors use to support students, Advisor 3 \textbf{shows more diversity in the topics covered across different meetings with students, suggesting that their meetings are more tailored to students' unique circumstances.} We measure across-meeting diversity by taking the average Jaccard Similarity Index (JSI) between codes applied to all possible pairs of each advisor's meeting comments, with a higher average JSI indicating more similarity between that advisor's meeting comments.\footnote{
For a pair of meeting comments, the JSI is defined as the ratio of the intersection to the union of the codes applied to each meeting. The maximum JSI of 1 indicates entirely overlapping codes, while the minimum JSI of 0 indicates entirely disjoint codes. We exclude comments with no codes.} Table \ref{tab:jaccard_codes} shows Advisor 3 has a notably lower average JSI.

Together, \textbf{these findings suggest the potential importance of \textit{advisor style} in effectively targeting student support,} illuminating an interesting direction for future research. It is also important to note the potential for unmeasured effects of advising style, such as student satisfaction or self-efficacy.

\section{Discussion}\label{sec:discussion}

Our study of the MAAPS intervention shines a light on the central role of advisor discretion in algorithm-supported advising. Combining quantitative and qualitative analysis, we provide diverse, suggestive evidence that discretionary action by advisors helps students achieve better educational outcomes. In this section, we discuss four implications of this finding for the design, evaluation, implementation, and normative aspects of algorithmic decision support, in college advising and beyond:
\begin{itemize}[itemsep=-1ex]
\item Algorithmic decision aids should support, not replace, advisors' discretionary judgment (\S\ref{sec:discussion:design}).
\item Evaluations of algorithmic decision aids that provide insight into how human discretion affects decision making ``in the wild'' are needed (\S\ref{sec:discussion:evaluation}).
\item  Where human discretion is important to an algorithmic decision aid's success, impacts of deployment may fail to generalize or scale (\S\ref{sec:discussion:implementation}).
\item It is important to consider both the practical and normative roles of human discretion before automating decision making (\S\ref{sec:discussion:normative}).
\end{itemize}

\subsection{Designing Algorithmic Aids for Human Expertise}\label{sec:discussion:design}
Much  of the research on the design of algorithmic decision aids ignores the role of human decision makers, and a recent literature review finds that algorithm design in higher education is no exception to this trend \cite{mcconvey_human-centered_2023}. Our work suggests this is a missed opportunity. When the exercise of human discretion can improve student outcomes, algorithmic aids can be made more effective by being explicitly designed to support and enhance, rather than replace, human expertise \cite{liuReimaginingMachineLearning2023,holsteinConceptualFrameworkHuman2020}. 

There are many ways to incorporate human discretion in the design of algorithmic tools. In addition to using human-centered design principles \cite{mcconvey_human-centered_2023}, algorithm designers could draw on the literature on complementarity to optimize for joint human-algorithmic performance \cite{donahue_human-algorithm_2022}. New advisors could be trained by large language models with embedded expert strategies \cite{wang-etal-2024-bridging}. Algorithmic tools could be used to automate routine administrative actions that do not require expertise, freeing advisors’ time to focus on more difficult tasks \cite{snyder2024algorithm}. And algorithms can be used to suggest ``tips'' that address bottlenecks in advisor decision making \cite{bastaniImprovingHumanSequential2025}.

Our study also identifies significant disadvantages to the increasingly frequent proposal of replacing human advisors with automation. In particular, it is difficult to capture 
all of the non-algorithmic context used in college advising through increased data collection or feature extraction from unstructured data (as in e.g., \cite{vallonPatientLevelClinicalExpertise2022}). Much of the non-algorithmic information used by MAAPS advisors, such as the psychological constructs of student engagement and motivation, is difficult to encode quantitatively \cite{fredricksStudentEngagementContext2016,fredricksjenniferAddressingChallengeMeasuring2019}. Advisors also use non-algorithmic context to address data gaps, such as missing course credit information or delays in placement test results, echoing a pattern in other domains \cite{chengHowChildWelfare2022}. While chatbots could gather up-to-date information in principle, \emph{interpersonal relationships} and \emph{trust} are essential to  obtaining information about sensitive topics and for providing meaningful support \cite{rastogiTaxonomyHumanML2023}. Finally, expanded structured data collection about students can act as a form of surveillance \cite{citron_surveilled_2024}, with risks like “anticipatory conformity,” students altering their behavior to avoid being algorithmically flagged \cite{mcconveyThisNotData2024}. 

\subsection{Evaluating Discretion in the Wild}\label{sec:discussion:evaluation}
The importance of human discretion to the MAAPS experiment also has implications for evaluating decision support systems. Our results highlight the need for developing experimental methodology that allows us to understand the role of discretion in the success of algorithmic aids. 

Some researchers have attempted to study human discretion through controlled experiments with crowdworkers \cite{greenDisparateInteractionsAlgorithmintheLoop2019,bucincaTrustThinkCognitive2021,poursabzi-sangdehManipulatingMeasuringModel2021,grgic-hlacaHumanDecisionMaking2019}, but such studies often lack external validity, failing to capture the knowledge of real decision makers and the complexity of institutional contexts \cite{lurie2020crowdworkers,hullman2024decision}. Thus, in experimental evaluations of real-world algorithm deployments \cite{imaiExperimentalEvaluationAlgorithmassisted2023}, it is important to develop techniques that build a thorough understanding of how the human decision maker interacts with the decision aid, and how that interaction affects outcomes \cite{rajiEvaluatingPredictionbasedInterventions2025}. For example, experimenters can treat the human decision maker, non-algorithmic external context, and the algorithmic aid as study conditions, experimentally manipulating these conditions in order to disentangle how they interact. When manipulation is impractical, field experiments can still collect structured or qualitative data to build an understanding of this interaction.

Many of the limitations of the present work offer additional future directions for developing the methodology of studying discretion in the wild. One limitation of our framework is its focus on single-stage decision making, while collegiate advising (as in the MAAPS experiment) is a multistage process. An interesting future direction might be to generalize the framework to long-term outcomes  and timeseries data about students, and design studies that can identify long-range effects. Another limitation of our study is that the MAAPS dataset lacks risk score records and descriptions of how advisors used the algorithm. Future studies could collect these data to enable a situated understanding of how algorithmic tools are used.

\subsection{Implementation Challenges: Generalization and Scaling}\label{sec:discussion:implementation}

Even when algorithm-assisted interventions are designed and evaluated with human decision makers in mind, they face two challenges common to many educational interventions: generalizing the intervention to new populations and scaling it across institutional contexts \cite{kraftInterpretingEffectSizes2020,slavinRelationshipSampleSizes2009, jepsen2009class, list2022voltage}. When discretion plays a role in intervention outcomes, the success of an intervention also depends on the success of the human decision maker, which may not generalize or scale.

For example, when generalizing interventions to new institutional settings, human decision makers may face structural barriers to their success. The MAAPS study exemplifies this pattern: Unlike GSU, other participating institutions in the MAAPS study %
did not implement core components of the intervention, such as assigning students to primary advisors from a centralized advising office rather than across academic departments \cite{alamuddinMonitoringAdvisingAnalytics2018}.  No other institutions achieved the same positive treatment effects as GSU, which the study authors partly attribute to this implementation challenge \cite{rossmanMAAPSAdvisingExperiment2023}. 

Additionally, the effectiveness of decision aids may depend on the institutional ability to support and retain human advisors, posing a scaling challenge. Prior work has shown that increased capacity in human outreach programs leads to larger effect sizes and higher power in RCTs \cite{boutilier2024randomized}. 
This effect may have contributed to the success of the program at GSU, which has invested significantly in expanding its advising staff and, by the time of the experiment, the student-to-advisor ratio had dropped from 700:1 in 2011 \cite{rogers2019advisement} to  239:1, with a 150:1 ratio in the treatment arm as the MAAPS protocol requires \cite{rossmanMAAPSAdvisingExperiment2023}.
The lowered ratio may be critical to giving advisors the necessary time to consider non-algorithmic context. By contrast, many institutions in the MAAPS experiment did not maintain a 150:1 ratio in the treatment arm due to retention problems, another factor study authors cite for the lack of positive impact at these schools \cite{rossmanMAAPSAdvisingExperiment2023}. It is possible that schools adopting algorithmic systems as cost-saving measures to avoid hiring more advisors---placing existing advisors in resource-constrained situations where they feel pressured to follow algorithmic recommendations \cite{kawakamiImprovingHumanAIPartnerships2022} and cannot develop the skills needed to provide discretionary support---would fail to replicate positive outcomes.%

\subsection{The Practical and Normative Role of Discretion}\label{sec:discussion:normative}Our finding that discretion plays a positive role in algorithm-assisted college advising contributes to two long-standing debates about algorithmic decision making: First, does human discretion help or hinder the delivery of high-quality interventions? And second, when should its institutional role be increased, maintained or diminished in light of automated decision systems?

The empirical evidence on the first question is split \cite{vaccaroWhenCombinationsHumans2024}. On the positive side, discretion can enable flexible, “street-level” judgments by human decision makers, allowing them to override algorithmic recommendations based on real-time information and evolving goals \cite{alkhatib_street-level_2019, holsteinConceptualFrameworkHuman2020, kawakamiImprovingHumanAIPartnerships2022, saxenaFrameworkHighStakesAlgorithmic2021}. Discretion can also promote equity; for example, in child maltreatment call screening, call workers have helped close racial disparities by adjusting how they respond to algorithmic risk scores to counteract systemic biases in the data \cite{chengHowChildWelfare2022, gerchickDevilDetailsInterrogating2023}. In addition to these benefits, our study suggests that advisor discretion acts as a \emph{bridge} between predictions (e.g., risk scores) and interventions \cite{liuActionabilityOutcomePrediction2023}. By responding to non-algorithmic information from students, advisors  deliver meaningful interventions that go beyond the capabilities of an algorithm. Whether advisor discretion also provides additional equity benefits---such as counterbalancing systemic biases of risk scores---remains an important question for future research. 

On the negative side, discretion can hinder decision making, particularly in settings where decision quality is thought to depend mostly on prediction accuracy \cite{kleinbergPredictionPolicyProblems2015} (though this is not our setting), as algorithms often outperform human forecasters \cite{dawesClinicalActuarialJudgment1989}. Additionally, while human decision makers may have access to information beyond what the algorithm does~\cite{alurAuditingHumanExpertise2023},  people may misjudge when to override or defer to algorithmic advice, leading to over-reliance or under-reliance~\cite{balakrishnanHumanAlgorithmCollaboration2025}. Moreover, discretion can introduce bias, as demonstrated by the case of a pretrial risk assessment tool in Kentucky, where judges were more likely to override low-risk scores for Black defendants than for white ones \cite{albright}.

Ultimately, decisions about discretion cannot  only be about effectiveness but also must incorporate institutional values. The use of proprietary, opaque algorithms without a human in the loop can create accountability gaps, limiting the ability of students or advisors to contest decisions \cite{brown_toward_2019}. It has been shown that rigid institutional mandates to follow algorithmic recommendations can undermine worker autonomy, generate resentment, and shift the balance of power away from human workers  \cite{saxenaFrameworkHighStakesAlgorithmic2021, brayneTechnologiesCrimePrediction2021}.  Those classified as ``high-risk'' by risk prediction tools may face stigmatization  \cite{townsendWereNotLiving2023}, especially when predictions use demographic variables. Finally, students may prefer human discretion over algorithmic decision making \cite{johnsonPredictiveAlgorithmsPerceptions}, trusting trust human actors over algorithmic systems to prioritize their best interests \cite{dasWhyStudentsReject2025}. Particularly in educational contexts like advising, where personal connection and trust are critical, designing algorithmic systems must involve careful reflection on when and how to preserve human discretion.

\paragraph{Acknowledgments.} We would like to thank the hard work of the entire MAAPS project team at GSU. The efforts of the MAAPS advisors, particularly the diligence and dedication of Emily Buis, were instrumental in developing a rich source of data that made this work possible. We thank Timothy Renick for his visionary leadership at NISS and for sharing insights that provided valuable context and inspiration for this research collaboration. We also thank Tim Fulton, a prior staff member at NISS, for his support, as well as Andy Zhang for research assistantship in the preliminary stage of this study. We thank Manish Raghavan for valuable discussions on his earlier work.

Research reported in this manuscript was supported by an Amazon Research Award Fall 2023 (L.T.L. and K.S.). Any opinions, findings, and conclusions or recommendations expressed in this material are those of the authors and do not reflect the views of Amazon.

\newpage
\bibliographystyle{abbrvnat}
\bibliography{advisors}

\newpage
\appendix

\section{Appendix}
\subsection{MAAPS experiment and data}
\label{appendix:dataset}

\subsubsection{Baseline Characteristics}
\label{appendix:dataset:baseline}

For each student, the dataset provides values of the following variables at time of study enrollment:
\begin{itemize}
\item Student ID
\item Month and year of birth
\item Gender (selected from one of the following options: female, male, other)
\item Race/ethnicity (selected from one of the following options: non-resident alien, Hispanic of any race, American Indian or Alaska Native, Asian, Black or African-American, Native Hawaiian or Other Pacific Islander, White, Two or More Races)
\item Whether the student is a transfer student
\item Whether the student received a high school diploma or equivalent
\item The student's high school GPA
\item The student's SAT score, both composite and broken down by verbal and math sections
\item The student's ACT score, both composite and on the English and Math sections
\item Number of college credit hours  already earned by the student (for instance through AP classes)
\item The student's expected financial contribution for the 2016-2017 academic year (in twelve $\$1050$ intervals, ranging from $\$0-\$1050$ to $\$22,051$ or more)
\end{itemize}

\subsubsection{Timeseries Data}
\label{appendix:dataset:timeseries}

For each term, the dataset provides the value of the following variables for all students:

\begin{itemize}
\item Gender
\item Graduation term and year (if relevant)
\item Term enrollment status
\item Term GPA
\item Cumulative GPA
\item Declared majors and minors
\item Hours of credit attempted and earned in the term, broken down by remedial and non-remedial credits
\item Total number of non-remedial credits earned up to and including the term
\item Student residency in university-owned housing during the term
\item Student participation in another intensive advising program
\item Student residency status for tuition purposes
\item Student's assigned MAAPS advisor
\end{itemize}

\subsubsection{Meeting Logs}
\label{appendix:dataset:meeting_logs}
For students in the treatment arm, MAAPS advisors log every student interaction with the following information:
\begin{itemize}
\item What initiated the advising interaction with the student (one option selected from: registration or program review, advising system alert, regularly scheduled advising, early alert or instructor initiated, and student initiated)
\item Method of contact (one option selected from: in person, email, telephone, group, online, text, student's primary/departmental advisor, other)
\item Intervention(s) performed in the meeting (any number selected from a list of twenty, described below)
\item Freeform comments on the meeting
\end{itemize}

The list of twenty interventions were as follows, with descriptions from the MAAPS implementation guide provided to institutions participating in the study:
\begin{enumerate}
    \item Advised to seek academic support: The student is encouraged to seek academic support hrough tutoring or supplemental instruction. The purpose of the intervention is to help the student earn a grade that will support their academic progression and prevent them from failing or dropping a course.
    \item Advised to correct schedule: The student is encouraged to seek academic support through tutoring or supplemental instruction. The purpose of the intervention is to help the student earn a grade that will support their academic progression and prevent them from failing or dropping a course.
    \item Advised to declare or change major: Student is encouraged to declare or change their major based on advisor recommendations (potentially based on predictive analytics). Purpose of the intervention is to ensure student is able to complete the academic markers needed to fulfill all degree requirements in a timely manner.
    \item Advised about progression issues: The purpose of the intervention is to encourage the student to address curricular progression issues. Examples of such issues include enrollment in low number of hours, dropping classes, withdrawing from classes, receiving incompletes for coursework and other issues.
    \item Advised to adjust course balance: The purpose of the intervention is to ensure that the student takes the appropriate course balance and does not take a disproportionate number of high risk courses within the same term. The appropriate course balance information may be based on predictive analytics or the grades that the student has earned in prerequisite or gateway courses.
    \item Advised to see instructor: The student is encouraged to seek the advice/assistance of their course instructor in response to early alert notification for poor grades, lack of attendance or other academic concerns.
    \item Advised to seek academic coaching: The student is encouraged to seek specialized assistance related to non-cognitive academic skills including time management, note taking, reading text books, motivation etc.
    \item Advised to seek departmental or faculty guidance:  The purpose of the intervention is to encourage the student to build relationships with faculty in order to better understand professional, career or graduate school opportunities related to the student’s academic program. Use this category when the student is advised to see the faculty for reasons other than academic performance in the classroom. Please use category 6 to record advised to see faculty based on academic concerns.
    \item Advised to address financial issues: The purpose of the intervention is to encourage the student to address financial issues that undermine retention or academic progression. This would include issues related to FAFSA and FAFSA completion, Student Account balances, scholarships etc.
    \item Advised to address non-financial collegiate requirements: Purpose of intervention is to encourage student to address non-financial collegiate requirements that undermine retention or academic progression. Requirements may include testing and placement, document submission and maybe related to: vaccinations and health, final transcripts, housing information etc.
    \item Advised to seek career counseling and service: The purpose of intervention is to encourage student to seek advice about how to develop career and professional plan.
    \item Advised to seek engagement opportunities on campus: The purpose of intervention is to encourage the student to get involved in co-curricular, recreation or wellness oriented activities.
    \item Advised to see professional counselor: The purpose of intervention is to encourage the student to see a professional counselor for non-academic personal issues.
    \item Advised to seek disability services on campus: The purpose of intervention is to encourage the student to seek support of campus disability services because of learning disability or physical handicap.
    \item Advised to see medical professional: The purpose of intervention is to encourage the student to see a medical professional in response to health or wellness-related concerns.
    \item Advised to schedule an appointment: The purpose of intervention is to encourage the student to schedule a follow-up appointment with their MAAPS advisor.
    \item Advised about recommended schedule: The purpose of the intervention is to recommend courses that student should take in the future. This intervention should be recorded when the MAAPS advisor provides a recommended schedule. If the MAAPS advisors correct a schedule, the intervention should be recorded as 2, 4 or 5 above.
    \item Advised to see primary program advisor: The purpose of the intervention is to encourage the student to meet with a program advisor familiar with academic and non-academic program requirements and opportunities including internships, seminars, and other activities related to the student’s program of study.
    \item Other (Describe in comments)
    \item Advised through the in-person development or review of the academic map/degree planner: This category should be indicated if an advisor works 1-on-1 with the student to develop or review an academic map or planner.
\end{enumerate}

\subsubsection{Data Quality}\label{app:data:quality}
\begin{figure}[ht]
    \centering
\includegraphics[width=1\textwidth]{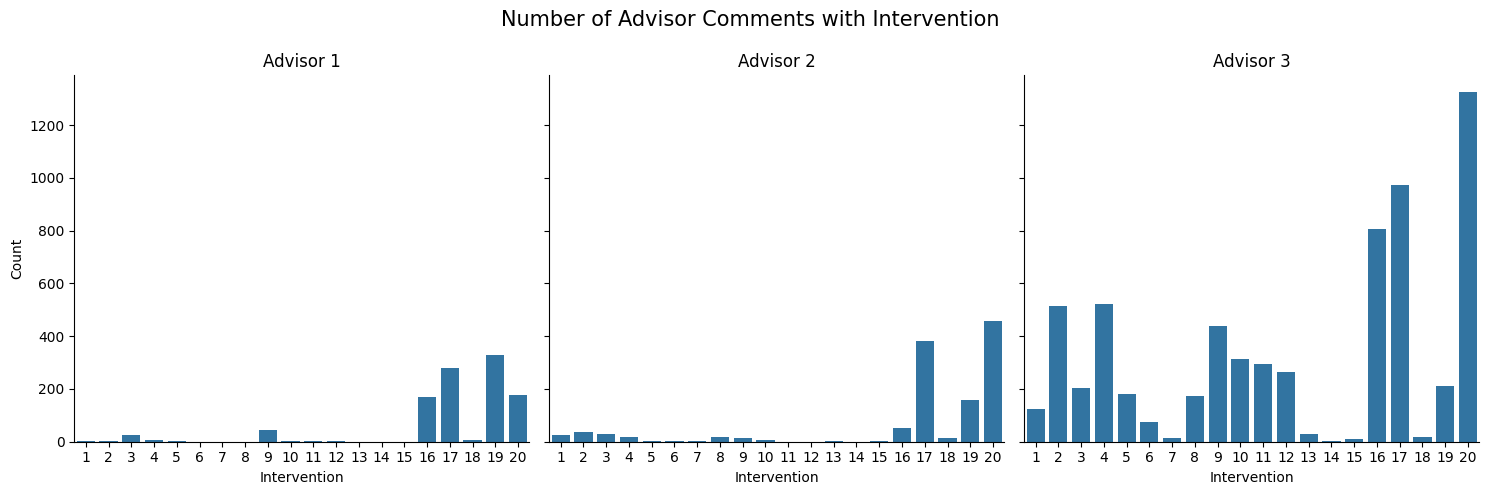}
    \caption{\normalfont Histograms of applied interventions per advisor, demonstrating an especially low diversity of interventions applied by Advisor 1. Advisors 4 and 5 are excluded  due to the limited number of available comments for each.}
    \label{fig:int_adv_distrib} 
\end{figure}

Our exploratory qualitative analysis reveals some inconsistency in the way that one advisor, Advisor 1, documented the interventions they performed during meetings. While sometimes there is clear evidence in the free-form comments that Advisor 1 performed a certain intervention during an interaction, they failed to mark the corresponding intervention boxes on the provided checklist. Additionally, Advisor 1 only indicated they had applied 10 of the 20 total interventions and only indicated they had applied four of them more than 100 times, even when their comments indicated that they had applied many additional interventions. We show how the distribution of interventions applied varies for each advisor in Figure \ref{fig:int_adv_distrib}. To account for reduced intervention data quality, we repeat the $\text{ExpertTest}^{*}$ excluding Advisor 1 to establish robustness of the results.

\subsection{Theoretical Framework}

\subsubsection{Proof of Proposition \ref{prop:nonidentifiability}}\label{app:theory:three_scms}

\begin{proof}
We prove Proposition \ref{prop:nonidentifiability} by defining two SCMs $\mathcal{M}_1$ and $\mathcal{M}_2$ such that both have the same joint distribution over $(A_t, Y_{t+1})$ but only $\mathcal{M}_1$ meets the definition for expertly targeted action. Without loss of generality, we construct $\mathcal{M}_1$ and $\mathcal{M}_2$ such that there is no algorithmically available context $X_t$, corresponding to a setting where the algorithmic recommendations are ignored by the human decision-maker.\footnote{We note the example can be modified to include $X_t$, for example by distributing it as a Bernoulli and adding a term to the equations for $Y_{t+1}$ and $A_t$ multiplying the entire expression by $X_t$ in both models.}

 Let $\textsf{Ber}(p)$ denote a Bernoulli random variable with mean $p$ and let $p\in (0,1)$. Consider the following two SCMs:
 
\begin{minipage}
{0.45\linewidth}
\centering
\underline{$\mathcal{M}_1$}
\begin{align*}
\N_{A_t} &\sim \textsf{Ber}(p) \\
\N_{U_t} &\sim \textsf{Ber}(p) \\
\N_{Y_{t+1}} &\sim \textsf{Ber}(p) \\
U_t &:= \N_{U_t} \\
A_t &:= \max(U_t,\N_A) \\
Y_{t+1} &:= \min(A_t,\max(U_t,\N_{Y_{t+1}}))
\end{align*}
\end{minipage}
\hfill
\begin{minipage}{0.45\linewidth}
\centering
\underline{$\mathcal{M}_2$}
\begin{align*}
\N_{A_t} &\sim \textsf{Ber}(2p-p^2) \\
\N_{Y_{t+1}} &\sim \textsf{Ber}(\frac{p^2-p-1}{p-2}) \\
A_t & := \N_A \\
Y_{t+1} & := \min(\N_{Y_{t+1}},A_t)
\end{align*}
\end{minipage}

\vspace{2mm}
First, we show that $\mathcal{M}_1$ and $\mathcal{M}_2$ are observationally equivalent. To derive the observational distribution  $\mathbb{P}(A_t , Y_{t+1})$ for each model, we sum the probabilities of all exogenous variable configurations that result in each setting of $A_t$ and $Y_{t+1}$. Table \ref{tab:noise_tables} demonstrates how the exogenous noise settings lead to different settings over model variables $\mathbf{V}$. Table \ref{tab:obs_dist} presents the observational distribution over $A_t$ and $Y_{t+1}$ for both models. Since the results are identical, this shows the models are observationally equivalent with respect to $\mathbb{P}(A_t, Y_{t+1})$.

\begin{table}[ht]
\centering
\begin{minipage}{0.48\textwidth}
\centering
\textbf{Noise table for $\mathcal{M}_1$} \\[0.5em]
\begin{footnotesize}
\begin{tabular}{c|c|c|c|c|c|c}
$\N_{A_t}$ & $\N_{U_t}$ & $\N_{Y_{t+1}}$ & $U_t$ & $A_t$ & $Y_{t+1}$ & Probability\\
\hline
0 & 0 & 0 & 0 & 0 & 0 & $-p^3+3p^2-3p+1$\\
0 & 0 & 1 & 0 & 0 & 0 & $p^3-2p^2+p$\\
0 & 1 & 0 & 1 & 1 & 1 & $p^3-2p^2+p$ \\
0 & 1 & 1 & 1 & 1 & 1 & $p^2-p^3$\\
1 & 0 & 0 & 0 & 1 & 0 & $p^3-2p^2+p$\\
1 & 0 & 1 & 0 & 1 & 1 & $p^2-p^3$\\
1 & 1 & 0 & 1 & 1 & 1 & $p^2-p^3$\\
1 & 1 & 1 & 1 & 1 & 1 & $p^3$
\end{tabular}
\end{footnotesize}
\end{minipage}
\hfill
\begin{minipage}{0.48\textwidth}
\centering
\textbf{Noise table for $\mathcal{M}_2$} \\[0.5em]
\begin{footnotesize}
\begin{tabular}{c|c|c|c|c}
$\N_{A_t}$ & $\N_{Y_{t+1}}$ & $A_t$ & $Y_{t+1}$ & 
Probability\\
\hline
0 & 0 & 0 & 0 & $\frac{-p^4+4p^3-6p^2+4p-1}{p-2}$\\
0 & 1 & 0 & 0 & $\frac{p^4-3p^3+2p^2+p-1}{p-2}$\\
1 & 0 & 1 & 0 & $p^3-2p^2+1$\\
1 & 1 & 1 & 1 & $-p^3+p^2+p$
\end{tabular}
\end{footnotesize}
\end{minipage}
\caption{\normalfont Noise tables for models $\mathcal{M}_1$ and $\mathcal{M}_2$. These tables show the values of the endogenous variables under different exogenous noise settings and their associated probabilities.}
\label{tab:noise_tables}
\end{table}

\begin{table}[ht]
\centering
\begin{tabular}{c|c|c}
$A_t$ & $Y_{t+1}$ & $\mathbb{P}(A_t,Y_{t+1})$ \\
\hline
0 & 0 & $p^2-2p+1$\\
0 & 1 & $0$\\
1 & 0 & $p^3-2p^2+p$\\
1 & 1 & $-p^3+p^2+p$
\end{tabular}
\caption{\normalfont Observational distribution over $A_t$ and $Y_{t+1}$ in both $\mathcal{M}_1$ and $\mathcal{M}_2$, obtained by summing over the rows of the noise table that produce these settings of $A_t$ and $Y_{t+1}$ for each model.}\label{tab:obs_dist}
\end{table}

Next, we show both of these models meet Assumptions \ref{assumption:dag} through \ref{assumption:faithfulness}. By construction, both models meet Assumption \ref{assumption:dag}, inducing DAGs meeting the graphical constraints, and Assumption \ref{assumption:markovity}, as noise variables are distributed independently. We now show that both $\mathcal{M}_1$ and $\mathcal{M}_2$ meet Assumption \ref{assumption:faithfulness} of faithfulness, from which causal minimality (Assumption \ref{assumption:minimality}) follows. Demonstrating faithfulness requires showing that, for all variables, a lack of $d$-separation implies statistical dependence. We show this is true for $\mathcal{M}_1$ below, with probabilities calculated using Table \ref{tab:noise_tables}.

\begin{enumerate}[label=(\arabic*)]
\item $U_t\not\perp A_t$

\begin{align*}
\mathbb{P}(U_t=0,A_t=1)&=p-p^2\\
\mathbb{P}(U_t=0)\times \mathbb{P}(A_t=1) &=p^3-3p^2+2p
\end{align*}

\noindent For all $p\in(0,1)$, $p-p^2\neq p^3-3p^2+2p$. Thus, $U_t\not\perp A_t$.

\item $U_t\not\perp Y_{t+1}$

\begin{align*}
\mathbb{P}(U_t=0,Y_{t+1}=1)&=p^2-p^3\\
\mathbb{P}(U_t=0)\times \mathbb{P}(Y_{t+1}=1)&=p^4-2p^3+p
\end{align*}

\noindent For all $p\in(0,1)$, $p^2-p^3\neq p^4-2p^3+p$. Thus, $U_t\not\perp Y_{t+1}$.

\item $A_t\not\perp Y_{t+1}$
\begin{align*}
\mathbb{P}(A_t=0,Y_{t+1}=1)&=0\\
\mathbb{P}(A_t=0)\times \mathbb{P}(Y_{t+1}=1)&=-p^5+3p^4-2p^3-p^2
\end{align*}
\noindent For all $p\in(0,1)$, $=-p^5+3p^4-2p^3-p^2\neq 0$. Thus, $A_t\not\perp Y_{t+1}$.
\item $U_t\not\perp A_t|Y_{t+1}$
\begin{align*}
\mathbb{P}(U_t=1,A_t=1|Y_{t+1}=1)&=\frac{p^2}{-p^2+p+1}\\
\mathbb{P}(U_t=1|Y_{t+1}=1)\times \mathbb{P}(A_t=1|Y_{t+1}=1)&=\frac{1}{-p^2+p+1}
\end{align*}
\noindent For all $p\in(0,1)$, $\frac{p^2}{-p^2+p+1}\neq \frac{1}{-p^2+p+1}$. Thus, $U_t\not\perp A_t|Y_{t+1}$.

\item $U_{t+1} \not \perp Y_{t+1} | A_t $
\begin{align*}
\mathbb{P}(Y_{t+1}=1,U_{t}=0|A_{t}=1) &=\frac{p^2-p^3}{2p-p^2}\\
\mathbb{P}(Y_{t+1}=1|A_t=1)\times \mathbb{P}(U_t=0|A_t=1)&=\frac{p^3 - 2 p^2 + 1}{p^2 - 4 p + 4}
\end{align*}

\noindent For all $p\in(0,1)$, $\frac{p^2-p}{p-2}\neq \frac{p^3 - 2 p^2 + 1}{p^2 - 4 p + 4}$. Thus, $Y_{t+1} \not \perp U_{t} | A_t $.

\item $A_{t} \not \perp Y_{t+1} | U_t$

\begin{align*}
\mathbb{P}(Y_{t+1}=1,A_t=1|U_t=0)&=p^2\\
\mathbb{P}(Y_{t+1}=1|U_t=0) \times \mathbb{P}(A_{t}=1|U_t=0)&=p^3
\end{align*}
\noindent For all $p\in(0,1)$, $p^2\neq p^3$. Thus, $Y_{t+1} \not \perp A_t | U_t$.
\end{enumerate}

To satisfy faithfulness, $\mathcal{M}_2$ must have $A_t\not\perp Y_{t+1}$. Since $\mathcal{M}_1$ is observationally equivalent to $\mathcal{M}_2$ over $(A_t, Y_{t+1})$, this follows from (3).

\quad Finally, we show $\mathcal{M}_1$ meets the three criteria of expertly targeted action (Definition \ref{def:expertly_targeted_action}) while  $\mathcal{M}_2$ does not.

We begin by showing $\mathcal{M}_1$ meets all three criteria:
 \begin{enumerate}
\item \textbf{Effective action:} 
\begin{align*}
\mathbb{P}(Y_{t+1}|do(A_t=1),U_t=1)&= \mathbb{P}(Y_{t+1}|A_t=1,U_t=1)&\text{Parents do/see$^*$}\\
&= 1&\text{Calculated from Table \ref{tab:noise_tables}}\\
&>0 \\
&=\mathbb{P}(Y_{t+1}|A_t=0,U_t=1)&\text{Calculated from Table \ref{tab:noise_tables}}\\
&=\mathbb{P}(Y_{t+1}|do(A_t=0),U_t=1)&\text{Parents do/see$^*$}
\end{align*}
\noindent $^{*}$ Parents do/see is that conditioning and intervening on parent values of a variable are interchangeable
 \citep[p.32]{bareinboimPearlsHierarchyFoundations2022}. It is true for all DAGs generated by SCMs \citep[p.33]{bareinboimPearlsHierarchyFoundations2022}.

\item \textbf{Targeted action:} 
\begin{align*}
\mathbb{P}(A_t=1|do(U_t=1)) & = \mathbb{P}(A_t=1|U_t=1)&\text{Parents do/see}\\
&=1&\text{Calculated from Table \ref{tab:noise_tables}}\\
&>p&p\in(0,1) \text{ by assumption }\\
&=\mathbb{P}(A_t=1|U_t=0)&\text{Calculated from Table \ref{tab:noise_tables}}\\
&=\mathbb{P}(A_t=1|do(U_t=0))&\text{Parents do/see}
\end{align*}

\item \textbf{Heterogeneous effects:}
\begin{align*}
\Delta(1)&=\mathbb{P}(Y_t=1|A_t=1,U_t=1)-\mathbb{P}(Y_t=1|A_t=0,U_t=1)&\text{Parents do/see}\\
&= 1 &\text{Calculated from Table \ref{tab:noise_tables}}\\
&> p & p\in(0,1)\\
&=\mathbb{P}(Y_t=1|A_t=1,U_t=0)-\mathbb{P}(Y_t=1|A_t=0,U_t=0))&\text{Calculated from Table \ref{tab:noise_tables}}\\
&=\Delta(0)& \text{Parents do/see}
\end{align*}
 \end{enumerate}

Meanwhile, $\mathcal{M}_2$ can meet neither of targeted action nor heterogeneous effects, since $U_t$ is not included in the model. Thus, $\mathcal{M}_1$ and $\mathcal{M}_2$ induce identical observational distributions $\mathbb{P}(A_t,Y_{t+1})$, yet only $\mathcal{M}_1$ exhibits expertly targeted action.
 \end{proof}

\begin{figure*}[ht]
\centering
\begin{minipage}{0.45\textwidth}
\centering
\begin{tikzpicture}[
    every node/.style={draw, circle, minimum size=1.2cm,scale=0.7},
    ->, thick, >=Stealth
]
\node (U) at (1, 0) {\(\mathbf{U}_t\)};
\node (X) at (1, 2) {\(\mathbf{X}_t\)};
\node (A) at (0, 1) {\(A_t\)};
\node (Y) at (2, 1) {\(Y_{t+1}\)};

\draw[dotted, ->] (X) -- (Y);
\draw[->] (U) -- (Y);
\draw[dotted, ->] (X) -- (A);

\draw[dotted, ->] (U) -- (A);
\draw[ ->,color=blue] (A) -- (Y);
\end{tikzpicture}
\caption*{\textcolor{blue}{(1) Impactful action}}
\end{minipage}
\hfill
\begin{minipage}{0.45\textwidth}
\centering
\begin{tikzpicture}[
    every node/.style={draw, circle, minimum size=1.2cm,scale=0.7},
    ->, thick, >=Stealth
]
\node (U) at (1, 0) {\(\mathbf{U}_t\)};
\node (X) at (1, 2) {\(\mathbf{X}_t\)};
\node (A) at (0, 1) {\(A_t\)};
\node (Y) at (2, 1) {\(Y_{t+1}\)};

\draw[dotted, ->] (X) -- (Y);
\draw[->] (U) -- (Y);
\draw[dotted, ->] (X) -- (A);

\draw[ ->,color=red] (U) -- (A);
\draw[dotted, ->] (A) -- (Y);
\end{tikzpicture}
\caption*{\textcolor{red}{(2) Non-algorithmic action}}
\end{minipage}
\hfill
\caption{\normalfont The two conditions of \textbf{\textcolor{blue}{impactful action}} and \textbf{\textcolor{red}{non-algorithmic action}} are implied by certain causal edges being present in the graph. The proof of Proposition \ref{prop:sufficiency} can thus use d-separation and Markovity to show these criteria are met.}\label{fig:graphical_model_expertise_intervention}
\end{figure*}

\subsection{Proof of Proposition \ref{prop:sufficiency}}\label{app:theory:sufficiency_proof}

\begin{proof}
By the Markov condition, if $A_t\not\perp Y_{t+1} |\ \mathbf{X}_t$ then  $A_t \not\perp_{\mathcal{G}_\mathcal{M}} Y_{t+1} |\ \mathbf{X}_t$. Thus, by the definition of $d$-separation, either  $A_t \to Y_{t+1}$ or $\mathbf{U}_t\to A_t$. First, consider the case that $A_t\rightarrow Y_{t+1}$. We will show that in this case, the impactful action criterion is met, which requires showing the following inequality holds:
\begin{align}\label{eq:impactful_action}
\mathbb{P}(Y_{t+1}=1|do(A_t=1),X_t=x,U_t=u) \neq 
\mathbb{P}(Y_{t+1}=1|do(A_t=0),X_t=x,U_t=u)
\end{align}

\noindent By Assumption \ref{assumption:minimality} of causal minimality, for any $Y$, for all $X\in \mathbf{PA}_Y$, 
$X\not\perp Y | PA_Y \backslash \{X\}$ \citep[p.109]{peters_elements_2017}. Since $A_t\in \mathbf{PA}_{Y_{t+1}}$ by assumption, for some $a$, $y$, $x$  and $u$, this implies:
\begin{small}
\begin{align}\label{eq:minimality}
\mathbb{P}(Y_{t+1}=y,A_t=a|X_t=x,U_t=u) \neq 
\mathbb{P}(Y_{t+1}=y|X_t=x,U_t=u) \times \mathbb{P}(A_t=a|X_t=x,U_t=u)
\end{align}
\end{small}
\noindent Using the product rule, we can manipulate the left hand side of Inequality~\ref{eq:minimality}:
\begin{small}
\[
P(Y_{t+1}=y,A_t=a|X_t=x,U_t=u) = P(Y_{t+1}=y|A_t=a,X_t=x,U_t=u)\times P(A_t=a|X_t=x,U_t=u)
\]
\end{small}
\noindent Then cancelling $P(A_t=a|X_t=x,U_t=u)$ from both sides of Inequality \ref{eq:minimality}, we get:
\begin{align}\label{eq:minimality-simplified}
P(Y_{t+1}=y|A_t=a,X_t=x,U_t=u) &\neq 
\mathbb{P}(Y_{t+1}=y|X_t=x,U_t=u)
\end{align}

\noindent Applying the law of total probability, we can now expand the right hand side of Inequality~\ref{eq:minimality-simplified} as:
\begin{align*}
\mathbb{P}(A_t=1|X_t=x,U_t=u) \times \mathbb{P}(Y_{t+1}=y|A_t=1,X_t=x, U_t=u) + \\
\mathbb{P}(A_t=0|X_t=x,U_t=u) \times \mathbb{P}(Y_{t+1}=y|A_t=0,X_t=x,U_t=u) 
\end{align*}

\noindent Without loss of generality, suppose $a=1$. Then by subtraction of $\mathbb{P}(A_{t}=1|X_t=x,U_t=u)\times \mathbb{P}(Y_{t+1}=y|A_t=1,X_t=x,U_t=u)$ from both sides of Inequality \ref{eq:minimality-simplified},

\begin{align*}
\mathbb{P}(A_t=0|X_t=x,U_t=u) \times \mathbb{P}(Y_{t+1}=y|A_t=1,X_t=x,U_t=u)\neq \\
\mathbb{P}(A_t=0|X_t=x,U_t=u) \times \mathbb{P}(Y_{t+1}=y|A_t=0,X_t=x,U_t=u)
\end{align*}

\noindent And so by cancellation of $\mathbb{P}(A_t=0|X_t=x,U_t=u)$ from both sides, 
\begin{align}
\mathbb{P}(Y_{t+1}=y|A_t=1,X_t=x,U_t=u) \neq 
\mathbb{P}(Y_{t+1}=y|A_t=0,X_t=x,U_t=u)
\end{align}

\noindent Since $Y_{t+1}$ is binary, we can replace $y$ with 1 (as if $y=0$, this inequality implies the same inequality holds for $Y_{t+1}=1$):
\begin{align}\label{eq:impactful_action_obs}
\mathbb{P}(Y_{t+1}=1|A_t=1,X_t=x,U_t=u) \neq 
\mathbb{P}(Y_{t+1}=1|A_t=0,X_t=x,U_t=u)
\end{align}

\noindent Finally, by applying parents do/see \citep[p.32] {bareinboimPearlsHierarchyFoundations2022} to $A_t$ in Inequality \ref{eq:impactful_action_obs}, we have that Inequality \ref{eq:impactful_action} is true, that is, impactful action is met. In the case that $\mathbf{U}_t\rightarrow A_{t}$, an analogous proof shows that non-algorithmic action is met. 
\end{proof}

\subsection{Generalization to Nonbinary Contrasts}\label{app:theory:nonbinary_contrasts}
Stating three causal conditions of expertly targeted action for nonbinary contrasts requires determining the importance of each causal contrast for each action and non-algorithmic information valuation. For instance, in determining whether a particular action taken by an advisor (say, recommending a change to a student's schedule in a meeting), was effective at improving a student's outcome (say, increasing their chances of graduation), what other action should we contrast it to? Should we contrast it to \textit{taking some other action} (say, recommending a student speak with their instructor)? If so, which one? Should we contrast it to taking \textit{no other action}, imagining that the advisor had not met with the student at all? Or should we combine these (and other) possibilities together somehow?

This is ultimately a normative choice with multiple defensible answers, depending on what alternate possibilities we deem relevant to attributing expertise. One way to formalize potential choices would be to stipulate weights on the relative importance of each causal contrast. For instance, to generalize effective action, for each $a\in R_{A_t}$, the support of $A_t$, define a collection of importance weights on the constrasts $W_a=\{w_{b} : b \in R_{A_t} /\ a \}$ such that $\sum W_a = 1$. A similar weighting can be defined for $\textbf{U}_t$, call it $W_u$. Then the causal contrasts to the action and non-algorithmic context evaluated in the components of the definition are weighted according to $W_a$ and $W_u$. For instance, effective action becomes checking that the particular action taken by the advisor, $a$, is more likely to have produced a good outcome than the alternative actions would have been, with the alternative actions weighted by $W_a$.

\subsection{Empirical Analysis of Advisor Expertise}
\label{appendix:cond_indep}

\subsubsection{Test Population}

\begin{figure}[!htb]
   \begin{minipage}{0.48\textwidth}
     \centering
     \includegraphics[width=.7\linewidth]{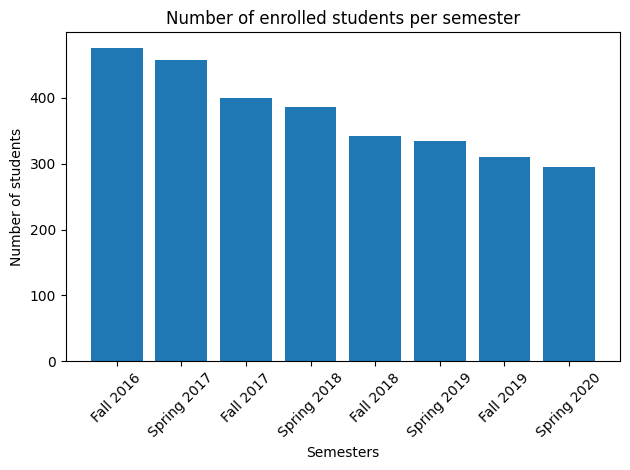}
     \caption{\normalfont Enrolled students per semester}\label{fig:enrolled_per_semester}
   \end{minipage}\hfill
   \begin{minipage}{0.48\textwidth}
     \centering
     \includegraphics[width=.7\linewidth]{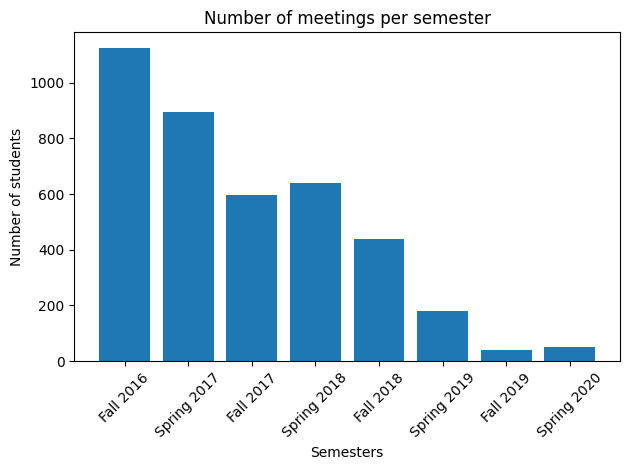}
     \caption{\normalfont Number of meetings per semester}\label{fig:meetings_per_semester}
   \end{minipage}
\end{figure}

Figure \ref{fig:enrolled_per_semester} visualizes the number of students enrolled each semester and Figure \ref{fig:meetings_per_semester} visualizes the number of meetings per semester.

\subsubsection{Positive vs. Negative Interventions}\label{appendix:cond_indep:test_methodology}

The ExpertTest* only rejects the null whenever the mean squared error in the synthetic datasets is routinely \textit{higher} than the the mean squared error of the original dataset. In \citet{alurAuditingHumanExpertise2023}, this one-sided test of predictions is sensible, as it ensures the test only detects instances where non-algorithmic information is used to improve the predictions of the human forecaster.

However, in the setting of intervention, we wish to detect actions that negatively predict outcomes, too, as these might indicate targeting of interventions to at-risk students. Thus we might be curious about whether actions $A_t$ predict $1-Y_i$, formally captured by  taking $MSE^{-}(\mathcal{D})=\sum_{i=1}^n (a_t^i - |1-y_{t+1}|)^2 = \sum_{i=1}^n  (|1-a_t^i|-y_{t+1})^2 $ (for binary $A_t^i,Y_{t+1}$). This corresponds to running two one-tailed tests on both directions of association.

Another approach would be to run a two-tailed test. Without loss of generality, suppose $\tau_K=\frac{J}{K}\leq 0.5$; then we would take the $p$-value as $\mathbb{P}(\tau_K\geq \frac{K-J}{K}\text{ or } \tau_K\leq \frac{J}{K})=2\times \frac{J+1}{K+1}$. While this test doubles uncorrected p-values, it halves the number of tested hypotheses; thus,  post-correction, the results are quite similar. For instance, tests of persistence yield the same results of just Intervention 17 (Recommend Schedule) as significant. For the GPA outcome, all the same interventions are significant except Intervention 2 (Correct Schedule), which drops just below significance ($p=0.059$).

We opt for the one-tailed test with transformed actions because it is simpler to interpret. The test of $A_t^+$ shows whether actions predict positive outcomes. The test of $A_t^{-}$ shows whether \textit{in}action predicts positive outcomes, equivalent to actions predicting negative outcomes. We note that the two-tailed test can be interpreted in a similar way, but it requires algebraically manipulating the test statistic.

\subsubsection{Test Versions and Parameters}
\label{appendix:cond_indep:test_parameters}

The ExpertTest* requires two parameters: $L$, the number of pairs, and $K$, the number of iterations. 

As mentioned in \S\ref{sec:expertise_quant:methodology}, we greedily pair $L$ ``approximately-identical'' students at baseline using a Euclidian distance function on fixed-scaled administrative features (excluding their race and ethnicity, which Georgia State University excludes from its predictors), prohibiting cross-semester matches for tests on semesterly outcomes.  We select the number of pairs $L=100$ for cumulative tests, $L=800$ for semesterly GPA tests, and $L=700$ for semesterly persistence tests, to yield an acceptable tradeoff of the number of pairs (which increases test power) with the quality of the matches (the more imperfect the matches, the higher the test's error).

\begin{figure}[ht]
\centering
\includegraphics[width=0.6\textwidth]{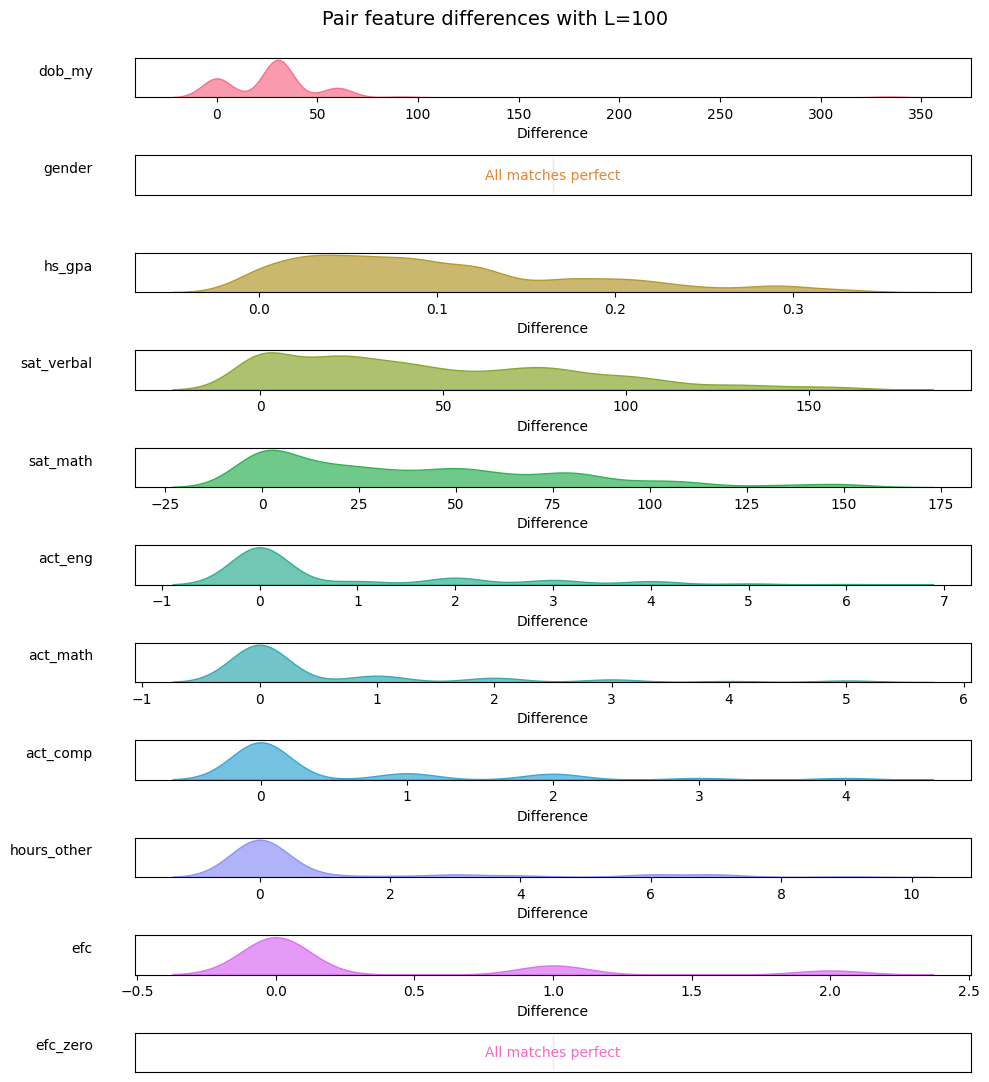}
\caption{\normalfont Differences between paired students' attributes (excluding race and ethnicity) at baseline for $L=100$ pairs. }\label{fig:euclidian_distance}
\end{figure}

For cumulative ExpertTest*, we visualize the differences by attribute between paired students in Figure \ref{fig:euclidian_distance} in the original units (e.g. date of birth in days, test scores in points).  The maximum  Euclidian distance between paired students in scaled feature space is $\approx 0.25$; when students are further away in one attribute, it is compensated for by a close match in other attributes. Pairs for the semesterly version of the test are slightly different (as there is a slightly different mix and smaller number of students enrolled each semester); Figure \ref{fig:euclidian_distance} is representative of match quality for these tests, with the highest Euclidian distance per semester between pairs rising at most to 0.49 in scaled feature space.

We set $K=1000$ to balance the trade-off between minimizing the ExpertTest* type I error---which decreases as $K$ increases \cite{alurAuditingHumanExpertise2023}---with computational constraints. The test's type I error includes an additive $\frac{1}{K+1}$ term, so the choice of $K = 1000$ ensures that this component contributes no more than $0.001$ to the test's error.

Additionally, we run the test on two timescales (early intervention and semesterly), and multiple different populations and outcome variables to ensure results are robust. 

Table \ref{table:test_summary} summarizes all our test parameters, variables and exclusions.

\begin{table}[ht]
\centering
 \resizebox{0.9\textwidth}{!}{
\begin{tabular}{c|c|c|c|c}
\hline
\textbf{Population} & $A_t$ & $\mathbf{Y}_{t+1}$ & $L$ & $K$ \\
\hline
Semesterly enrolled non-transfers*  &  Term intervention $i$ for $i\in[1,20]$ & Next-term persistence & 700 & 1000 \\
Semesterly enrolled non-transfers &  Term intervention $i$ for $i\in[1,20]$& Term GPA $\geq 2.75$ & 800 & 1000 \\
Semesterly enrolled non-transfers &  Term intervention $i$ for $i\in[1,20]$ & Term GPA $\geq 3.0$ & 800 & 1000 \\
Semesterly enrolled non-transfers &  Term intervention $i$ for $i\in[1,20]$ & Term GPA $\geq 3.25$ & 800 & 1000 \\
Semesterly enrolled non-transfers &  Term intervention $i$ for $i\in[1,20]$ & Term GPA $\geq 3.5$ & 800 & 1000 \\
Semesterly enrolled non-transfers &  Term intervention $i$ for $i\in[1,20]$ & Term GPA $\geq 3.75$ & 800 & 1000 \\
Semesterly enrolled non-transfers &  Term intervention $i$ for $i\in[1,20]$ & Term GPA $\geq 4.0$ & 800 & 1000 \\
Semesterly enrolled all transfer status* &  Term intervention $i$ for $i\in[1,20]$ & Next-term persistence & 700 & 1000 \\
Semesterly enrolled all transfer statuses &  Term intervention $i$ for $i\in[1,20]$ & GPA $\geq 3.5$ & 800 & 1000 \\
Semesterly enrolled non-transfer non-advisor 1*  &  Term intervention $i$ for $i\in[1,20]$ & Next-term persistence & 700 & 1000 \\
Semesterly enrolled non-transfer non-advisor 1 &  Term intervention $i$ for $i\in[1,20]$ & GPA $\geq 3.5$ & 800 & 1000 \\
Enrolled first year non-transfer &  First-year intervention $i$ for $i\in[1,20]$ & 4-year graduation & 100 & 1000 \\
Enrolled first year non-transfer &  First-year intervention $i$ for $i\in[1,20]$ & Cumulative GPA $\geq 3.5$ & 100 & 1000 
\end{tabular}}

\caption{\normalfont Versions of the ExpertTest* we run with different populations, $A_t$, $\mathbf{Y}_{t+1}$, $L$ (the number of pairs), and $K$ (the number of synthetic datasets). 
\\
*For tests on next-term persistence, we exclude enrolled students from spring 2020 in the population as the experiment concluded that term and no subsequent persistence data is available.}
\label{table:test_summary}
\end{table}

\subsubsection{Number of Swaps and Test Power}\label{appendix:quant_expertise:power_explanation}

Table \ref{table:swaps} shows the number of swaps between algorithmically-indistinguishable students that could increase or decrease the mean squared error (MSE) of synthetic datasets, for the semesterly tests on both persistence and GPA $\geq 3.5$ outcomes excluding potential transfer students. A smaller number of such swaps---whether increasing or decreasing the MSE---reduces the power of the test to detect conditional dependence. This is because when few swaps change the MSE, synthetic datasets are less likely to differ from the true dataset in MSE, even if non-algorithmic information is being used.

Thus, test power tends to decrease as the base rate of either the intervention or the outcome approaches 0 or 1. When interventions are rarely applied, most students will have the same intervention status, leaving little room for swaps to change the MSE. Similarly, because the base rate of persistence is relatively high (see Figure \ref{fig:enrolled_per_semester}), students are less likely to differ on this outcome, which reduces the test's power for persistence compared to GPA outcomes.

\begin{table}[ht]
\centering
 \resizebox{0.9\textwidth}{!}{\begin{tabular}{l|c|c|c|c|c}
\hline
\textbf{Intervention} & \textbf{Count} & 
\shortstack{\boldmath$\downarrow$ \textbf{MSE}\\ \textbf{GPA}} & 
\shortstack{\boldmath$\uparrow$ \textbf{MSE}\\ \textbf{GPA}} & 
\shortstack{\boldmath$\downarrow$ \textbf{MSE}\\ \textbf{Persistence}} & 
\shortstack{\boldmath$\uparrow$ \textbf{MSE}\\ \textbf{Persistence}} \\
\hline
1. Seek academic support & 162 & 19 & 11 & 11 & 5 \\
2. Correct schedule & 503 & 55 & 32 & 14 & 11 \\
3. Declare or change major & 238 & 23 & 18 & 8 & 4 \\
4. Progression issues & 666 & 63 & 35 & 20 & 12 \\
5. Adjust course balance & 192 & 22 & 19 & 11 & 10 \\
6. See instructor & 73 & 10 & 6 & 3 & 0 \\
7. Seek academic coaching & 14 & 1 & 0 & 0 & 0 \\
8. Seek departmental or faculty guidance & 190 & 15 & 23 & 6 & 3 \\
9. Address financial issues & 338 & 25 & 31 & 9 & 10 \\
10. Address non-financial collegiate requirements & 280 & 23 & 26 & 6 & 8 \\
11. Seek career counseling and service & 307 & 23 & 31 & 5 & 11 \\
12. Seek engagement opportunities on campus & 253 & 13 & 29 & 6 & 13 \\
13. See professional counselor & 28 & 1 & 4 & 2 & 2 \\
14. Seek disability services on campus & 2 & 0 & 0 & 0 & 0 \\
15. See medical professional & 9 & 1 & 0 & 0 & 0 \\
16. Schedule an appointment & 1035 & 76 & 45 & 25 & 13 \\
17. Recommended schedule & 1601 & 50 & 66 & 8 & 24 \\
18. See primary program advisor & 49 & 6 & 7 & 1 & 3 \\
19. Other (Describe in comments) & 635 & 7 & 19 & 14 & 6 \\
20. Develop degree planner & 1810 & 41 & 70 & 10 & 23
\end{tabular}}
\caption{\normalfont At $L=800$ pairs for GPA outcomes and $L=700$ for persistence outcomes, the above table records the total number of swaps of intervention $i$ for approximately algorithmically-indistinguishable pairs that could decrease $(\downarrow)$ vs. increase $(\uparrow)$ the mean squared error of the intervention at predicting the outcomes. Counts of the number of times the intervention was performed total are also included.}
\label{table:swaps}
\end{table}

\subsubsection{Full Results} \label{appendix:quant_expertise:results}

Results of the ExpertTest* are reported in Table \ref{table:term_notransfer_expertise_test_results} for term tests and Table \ref{table:early_intervention_notransfer_expertise_test_results} for early-intervention tests. Additionally, for term tests (which were the only tests with significant results), results including potential transfer students are reported in Table \ref{table:semester_intervention_transfers_expertise_test_results} and excluding students assigned Advisor 1 in Table \ref{table:semester_intervention_notransfer_noadvA_expertise_test_results}. Results are unchanged by exclusion of advisor 1, and remain similar upon inclusion of potential transfer students, which adds a new intervention, Intervention 8 (Seek Department or Faculty Guidance) as positively predictive of a high term GPA, and removes Intervention 2 (Correct Schedule) as negatively predictive of a low term GPA. The slight differences are likely reflective of slightly different characteristics of transfer student population and the inherent randomness of the test (small differences in the generated synthetic datasets).

Additionally, the top 5 most significant interventions at GPA binarization thresholds ranging from 2.5 to 4.0 are included in Table \ref{table:gpa_threshold_robustness}. As the table demonstrates, Interventions 4 (Progression Issues) and 16 (Schedule an Appointment) are significant for all tests, while Intervention 2 (Correct Schedule) and Intervention 20 (Develop Degree Planner) are always in the top 5 except for in predicting a perfect GPA of 4.0 (which would be expected due to the especially high bar in addition to the reduced power of the test when most people's outcomes do not differ).

\begin{table*}[h!]
\centering
 \resizebox{0.9\textwidth}{!}{
\begin{tabular}{r|c|c|c|c}
\hline 
\textbf{Intervention} & \textbf{Term GPA $(+)$} & \textbf{Term GPA $(+)$} & \textbf{Next-term persistence $(+)$} & \textbf{Next-term persistence $(-)$}\\
\hline
1&0.998002&0.399600&1.000000&0.399600
\\
2&0.998002&0.049950*&1.000000&0.882451\\
3&0.998002&0.633652&1.000000&0.607393
\\
4&0.998002&0.049950*&1.000000&0.607393\\
5&0.998002&0.777001&0.923692&0.882451
\\
6&0.998002&0.519481&1.000000&0.399600
\\
7&0.998002&0.639361&0.882451&0.882451
\\
8&0.444000&0.998002&1.000000&0.624830
\\
9&0.633652&0.998002&0.882451&0.923692
\\
10&0.777001&0.998002&0.882451&1.000000
\\
11&0.515848&0.998002&0.399600&1.000000\\
12&0.053280&0.998002&0.399600&1.000000
\\
13&0.455544&0.998002&0.882451&0.882451
\\
14&0.970458&0.961039&0.882451&0.882451\\
15&0.998002&0.639361&0.882451&0.882451
\\
16&0.998002&0.039960*&1.000000&0.333000
\\
17&0.399600&0.998002
&0.039960*&1.000000
\\
18&0.832851&0.998002&0.715951&1.000000
\\
19&0.053280&0.998002&1.000000&0.399600
\\
20&0.039960*&0.998002&0.119880&1.000000
\end{tabular}}
\caption{\normalfont Results of ExpertTest* on term outcomes excluding data from students missing data on transfer status, with Benjamini-Hochberg correction applied.}
\label{table:term_notransfer_expertise_test_results}
\end{table*}

\begin{table*}[h!]
\centering
 \resizebox{0.9\textwidth}{!}{
\begin{tabular}{r|c|c|c|c}
\hline 
\textbf{Intervention} & \textbf{Persistence $(+)$} & \textbf{Persistence $(-)$} & \textbf{Graduation $(+)$} & \textbf{Graduation $(-)$}\\
\hline
1& 0.866825 & 0.948143 & 0.943672 & 0.839161\\
2& 0.864590 & 0.948143 & 0.943672 & 0.839161\\
3& 0.864590 & 0.955801 & 0.839161 & 0.943672\\
4& 0.948143 & 0.880786 & 0.936563 & 0.839161\\
5& 0.866825 & 0.948143 & 0.943672 & 0.839161\\
6& 0.949051 & 0.864590 & 0.875791 & 0.875791\\
7& 0.880786 & 0.880786 & 0.875791 & 0.875791\\
8& 0.948143 & 0.880786 & 0.839161 & 0.943672\\
9& 0.864590 & 0.948143 & 0.839161 & 0.936563\\
10& 0.880786 & 0.948143 & 0.839161 & 0.943672\\
11& 0.346320 & 0.993007 & 0.839161 & 0.936563\\
12& 0.864590 & 0.949051 & 0.839161 & 0.936563\\
13& 0.880786 & 0.880786 & 0.936563 & 0.839161\\
14& 0.880786 & 0.880786 & 0.875791 & 0.875791\\
15& 0.948143 & 0.864590 & 0.875791 & 0.875791\\
16& 0.948143 & 0.864590 & 0.966034 & 0.839161\\
17& 0.239760 & 0.993007 & 0.839161 & 0.943672\\
18& 0.239760 & 0.993007 & 0.875791 & 0.936563\\
19& 0.880786 & 0.880786 & 0.936563 & 0.875791\\
20& 0.864590 & 0.949051 & 0.839161 & 0.936563

\end{tabular}}

\caption{\normalfont Results of ExpertTest* on early interventions for 4-year outcomes, excluding data from students missing data on transfer status, with Benjamini-Hochberg correction applied.}
\label{table:early_intervention_notransfer_expertise_test_results}
\end{table*}

\begin{table*}[h!]
\centering
 \resizebox{0.9\textwidth}{!}{
\begin{tabular}{r|c|c|c|c}
\hline 
\textbf{Intervention} & \textbf{Next-term persistence $(+)$} & \textbf{Next-term persistence $(-)$} & \textbf{Term GPA $(+)$} & \textbf{Term GPA $(-)$}\\
\hline
1& 1.000000 & 0.446220 & 1.000000 & 0.049950$^*$\\
2& 0.931925 & 0.882451 & 1.000000 & 0.164835\\
3& 0.882451 & 0.931925 & 1.000000 & 0.575425\\
4& 0.931925 & 0.882451 & 1.000000 & 0.204240\\
5& 0.931925 & 0.882451 & 1.000000 & 0.794994\\
6& 1.000000 & 0.693852 & 0.817183 & 1.000000\\
7& 0.882451 & 0.882451 & 1.000000 & 0.579421\\
8& 1.000000 & 0.446220 & 0.049950$^*$ & 1.000000\\
9& 0.446220 & 1.000000 & 0.399600 & 1.000000\\
10& 0.722611 & 1.000000 & 0.485668 & 1.000000\\
11& 0.446220 & 1.000000 & 0.399600 & 1.000000\\
12& 0.446220 & 1.000000 & 0.063936 & 1.000000\\
13& 0.882451 & 0.882451 & 0.399600 & 1.000000\\
14& 0.882451 & 0.882451 & 0.579421 & 1.000000\\
15& 0.882451 & 0.882451 & 1.000000 & 0.579421\\
16& 1.000000 & 0.259740 & 1.000000 & 0.039960$^*$\\
17& 0.019980$^*$ & 1.000000 & 0.108463 & 1.000000\\
18& 0.693852 & 1.000000 & 0.575425 & 1.000000\\
19& 1.000000 & 0.066600 & 0.093240 & 1.000000\\
20& 0.019980$^*$ & 1.000000 & 0.039960$^*$ & 1.000000

\end{tabular}}

\caption{\normalfont Results of ExpertTest* on semesterly interventions, including students with missing transfer status, with Benjamini-Hochberg correction applied.}
\label{table:semester_intervention_transfers_expertise_test_results}
\end{table*}

\begin{table*}[h!]
\centering
 \resizebox{0.9\textwidth}{!}{
\begin{tabular}{r|c|c|c|c}
\hline 
\textbf{Intervention} & \textbf{Next-term persistence $(+)$} & \textbf{Next-term persistence $(-)$} & \textbf{Term GPA $(+)$} & \textbf{Term GPA $(-)$}\\
\hline
1& 0.998002 & 0.365349 & 0.998002& 0.399600\\
2& 0.994903 & 0.775695 & 0.998002 & 0.049950$^*$\\
3& 0.994903 & 0.775695 & 0.998002 & 0.633652\\
4& 0.994903 & 0.365349 & 0.998002 & 0.049950$^*$\\
5& 0.994903 & 0.775695 & 0.998002 & 0.777001\\
6& 0.994903 & 0.775695 & 0.998002 & 0.519481\\
7& 0.920818 & 0.920818 & 0.998002 & 0.639361\\
8& 0.994903 & 0.365349 & 0.444000 & 0.998002\\
9& 0.994903 & 0.775695 & 0.633652 & 0.998002\\
10& 0.775695 & 0.994903 & 0.777001 & 0.998002\\
11& 0.775695 & 0.994903 & 0.515848 & 0.998002\\
12& 0.365349 & 0.994903 & 0.053280 & 0.998002\\
13& 0.994903 & 0.775695 & 0.455544 & 0.998002\\
14& 0.920818 & 0.920818 & 0.970458 & 0.961039\\
15& 0.920818 & 0.920818 & 0.998002 & 0.639361\\
16& 0.994903 & 0.365349& 0.998002 & 0.039960$^*$\\
17& 0.159840 & 0.998002 & 0.399600 & 0.998002\\
18& 0.775695 & 0.994903 & 0.832851 & 0.998002\\
19& 0.994903 & 0.775695 & 0.053280 & 0.998002\\
20& 0.365349 & 0.994903 & 0.039960$^*$ & 0.998002

\end{tabular}}
\caption{\normalfont Results of ExpertTest* on semesterly interventions, excluding data from students missing data on transfer status, and excluding Advisor 1, with Benjamini-Hochberg correction applied.}
\label{table:semester_intervention_notransfer_noadvA_expertise_test_results}
\end{table*}

\begin{table*}[h!]
\centering
 \resizebox{0.9\textwidth}{!}{
\begin{tabular}{l|l|l|l|l}
\hline 
 \textbf{Term GPA $\mathbf{\geq 2.75}$} & $\mathbf{\geq 3.0}$ & $\mathbf{\geq 3.25}$ &
$\mathbf{\geq 3.75}$ &
$\mathbf{\geq 4.0}$ \\
\hline
 4 (-) $p=0.019980^*$ & 4 (-) $p=0.019980^*$ & 4 (-) $p=0.019980^*$ & 4 (-) $p=0.019980^*$ & 4 (-) $p=0.019980^*$ \\ 
 16 (-) $p=0.019980^*$ & 16 (-) $p=0.019980^*$ &  16 (-) $p=0.019980^*$ &  16 (-) $p=0.019980^*$ &  16 (-) $p=0.019980^*$\\
 2 (-) $p=0.026640^*$ & 12 (+) $p=0.053280$  & 2 (-) $p=0.026640^*$ & 17 (+) $p=0.303696$ & 6 (-) $p=0.346320$\\
12 (+) $p=0.039960^*$ & 20 (+) $p=0.055944$ & 12 (+) $p=0.069930$ & 20 (+) $p=0.303696$ & 4 (-) $p=0.499500$ \\
20 (+) $p=0.446220$ & 2 (-) $p=0.055944$ & 9 (-) $0.139860$ & 2 (-) $p=0.303696$ & 19 (+) $0.511489$ \\
\end{tabular}}
\caption{\normalfont Top five most significant interventions on the ExpertTest* at GPA thresholds ranging from 2.75 to 4.0, and their $p$-values with Benjamini-Hochberg correction applied.}
\label{table:gpa_threshold_robustness}
\end{table*}

\begin{table*}[h!]
\centering
 \resizebox{0.9\textwidth}{!}{\begin{tabular}{l|c|c|c|c|c}
\hline
\textbf{Intervention} & \textbf{Count} & 
\shortstack{\boldmath$\downarrow$ \textbf{MSE}\\ \textbf{GPA}} & 
\shortstack{\boldmath$\uparrow$ \textbf{MSE}\\ \textbf{GPA}} & 
\shortstack{\boldmath$\downarrow$ \textbf{MSE}\\ \textbf{Persistence}} & 
\shortstack{\boldmath$\uparrow$ \textbf{MSE}\\ \textbf{Persistence}} \\
\hline
1. Seek academic support & 162 & 19 & 11 & 11 & 5 \\
2. Correct schedule & 503 & 55 & 32 & 14 & 11 \\
3. Declare or change major & 238 & 23 & 18 & 8 & 4 \\
4. Progression issues & 666 & 63 & 35 & 20 & 12 \\
5. Adjust course balance & 192 & 22 & 19 & 11 & 10 \\
6. See instructor & 73 & 10 & 6 & 3 & 0 \\
7. Seek academic coaching & 14 & 1 & 0 & 0 & 0 \\
8. Seek departmental or faculty guidance & 190 & 15 & 23 & 6 & 3 \\
9. Address financial issues & 338 & 25 & 31 & 9 & 10 \\
10. Address non-financial collegiate requirements & 280 & 23 & 26 & 6 & 8 \\
11. Seek career counseling and service & 307 & 23 & 31 & 5 & 11 \\
12. Seek engagement opportunities on campus & 253 & 13 & 29 & 6 & 13 \\
13. See professional counselor & 28 & 1 & 4 & 2 & 2 \\
14. Seek disability services on campus & 2 & 0 & 0 & 0 & 0 \\
15. See medical professional & 9 & 1 & 0 & 0 & 0 \\
16. Schedule an appointment & 1035 & 76 & 45 & 25 & 13 \\
17. Recommended schedule & 1601 & 50 & 66 & 8 & 24 \\
18. See primary program advisor & 49 & 6 & 7 & 1 & 3 \\
19. Other (describe in comments) & 635 & 7 & 19 & 14 & 6 \\
20. Academic map/degree planner & 1810 & 41 & 70 & 10 & 23
\end{tabular}}
\caption{\normalfont At $L=800$ pairs for GPA outcomes and $L=700$ for persistence outcomes, the above table records the total number of swaps of intervention $i$ for approximately algorithmically-indistinguishable pairs that could decrease $(\downarrow)$ vs. increase $(\uparrow)$ the mean squared error of the intervention at predicting the outcomes. Counts of the number of times the intervention was performed total are also included.}
\label{appendix:table:swaps}
\end{table*}

\subsubsection{Correlations Between Interventions}\label{app:rq1_quant:correlations} 
\begin{figure}[h] 

    \centering
\includegraphics[width=0.5\linewidth]{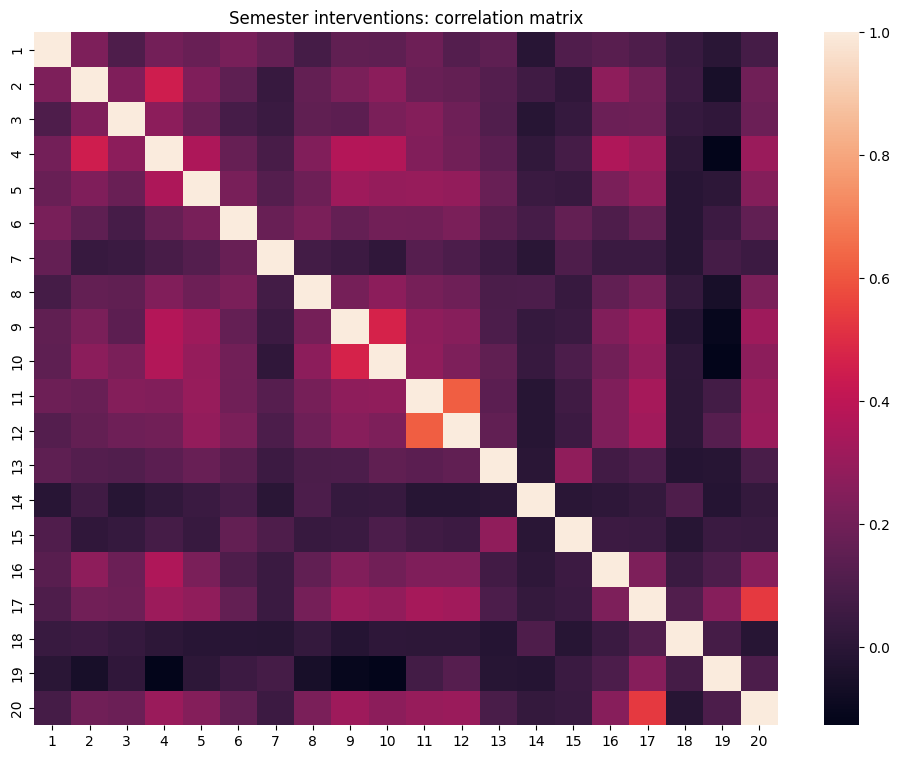} \caption{\normalfont Pearson correlation matrix between intervention per semester variables $A_t^i$ for $i\in[1,20]$.}
\label{fig:correlation_matrix}
\end{figure}

A significant result on the ExpertTest* might reflect dependence driven by application of interventions \textit{together}, since advisors may apply multiple interventions in one meeting. To address this possibility without having to test $2^{20}$ hypotheses (every possible combination of interventions), we here interpret results in light of the Pearson correlations between significant interventions, visualized in Figure \ref{fig:correlation_matrix}, on the assumption that any  interventions that may be significant due to joint application will be detected as separately significant.

For persistence outcomes, only one intervention (Intervention 17, Recommend Schedule) is significant, so no correlations warrant further attention. For GPA outcomes, Intervention 4 (progression issues) shows moderate correlation with each of the other significant interventions: Intervention 2 (Correct Schedule, $r = 0.44$, $p = 6.4 \times 10^{-117}$), Intervention 16 (Schedule Appointment, $r = 0.35$, $p = 1.31 \times 10^{-72}$), and Intervention 20 (Degree Map, $r = 0.30$, $p = 1.5 \times 10^{-53}$), while the remaining pairs of interventions show only weak correlation. Notably, Intervention 20 (Degree Map) predicts a GPA $\geq 3.5$ positively while Intervention 4 (Progression Issues) predicts it negatively, making it unlikely that their conditional dependence with outcomes reflects joint causal effects or targeting. Meanwhile, qualitative analysis supports the interpretation that Interventions 2 (Correct Schedule) and 16 (Schedule Appointment) are informed by different types of non-algorithmic information than Intervention 4 (Progression Issues) (Table~\ref{table:reason_counts}), suggesting they are separately targeted. However, given the moderate correlations, it remains possible that Interventions 2 (Correct Schedule) and 4 (Progression Issues), as well as Interventions 4 (Progression Issues) and 16 (Schedule Appointment), are either jointly non-algorithmically targeted or jointly influence GPA outcomes.

\section{Exploratory Qualitative Analysis}\label{app:qual_exploratory}

\subsection{Methods}\label{app:qual_exploratory:methods}

In the first stage of our qualitative analysis, we conduct exploratory coding on all 1,594 advisor meetings comments to understand the nature of advisor-student interactions as documented by those comments. In our analysis, codes served as succinct, essence-capturing labels that represent meaningful patterns in the data \cite{saldana2021coding}. 

\paragraph{Codes}

To determine the codes for our qualitative analysis, Authors 1 and 5 first coded subsamples of comments independently with fine-grained codes, and then met to design a codebook organizing all into three major thematic categories. Author 1 then coded the entire dataset for these three thematic categories using the \textit{Atlas.ti} software. Our complete codebook with an example quote for each category is provided below. 

\begin{itemize}
    \item \textbf{Academic Matters:} The advisor and student discuss an \textit{academic} topic of immediate relevance to their degree, including scheduling, academic administrative issues like degree requirements, major/minor selection, academic performance, or review of institutional benchmarks.
    \begin{quote}
        ``Student is really liking three of their four classes. MATH [course number] is giving them a big difficulty, and they know they cannot continue. They did do Summer Success\footnote{a program for incoming students to take summer credits}, so they will still be on track for a four year plan. Student likes summer, and planned to take math again in the summer. Student needs to talk to academic coach to release early alert grade.''        
    \end{quote}
    
    \item \textbf{Non-academic Matters:} The advisor and the student discuss a \textit{non-academic} topic. This category includes career aspirations, planning for graduate school, financial matters such as aid or non-payment of tuition, extracurriculars, and personal life.  While the discussion may touch on academics (for example, discussing academic context in the context of deciding a career or graduate school), it must go beyond the scope of immediate discussion of the student's course performance or plans in their undergraduate degree. 
    \begin{quote}
    ``Student did Summer Success, so they have made friends. But they are currently living with friends and family. [Redacted], because of such, do not have a permanent place to call home.''
    \end{quote}
    
    \item \textbf{Relationship Development:} The advisor demonstrates development of a \textit{relationship} or \textit{emotional connection} to the student. This can come in the form of expressing concern, understanding, or support for students, noting the student's emotional state during the interaction, or recording observations about the student's personality.
    \begin{quote}
    ``New freshman --- an introvert, so while they didn't have too much to say how their classes or GSU was going, I could tell subtly that they did like it.''
    \end{quote}
\end{itemize}

\subsection{Results}\label{app:qual_exploratory:results} To supplement our analysis of results in \S\ref{sec:expertise_qual:styles}, we report the frequency of codes by each advisor in Table \ref{tab:qual_codes_counts}.

\begin{table}
\centering
\begin{tabular}{l|ccc}
\hline
 & \textbf{Advisor 1} & \textbf{Advisor 2} & \textbf{Advisor 3} \\
\textbf{Code} & (851 comments) & (426 comments) & (375 comments)\\
\hline
Academic & 654 & 343 & 290\\
Non=academic & 153 & 63 & 135\\
Relationship & 63 & 17 & 101
\end{tabular}
\caption{\normalfont Frequencies of the applied codes by each advisor during student-advising meetings, based on our exploratory qualitative analysis. Code counts per advisor may sum to a different number than the total number of comments per advisor as meeting comments can have zero or multiple codes applied.}
\label{tab:qual_codes_counts}
\end{table}

\paragraph{Academic vs. holistic style} 

Here we expand on what we call the ``academic style'' of Advisor 1 and 2. There are two features of the academic style of advising. First, it tends to be very structured, especially for Advisor 1's comments, which often begin with a routinized list of advising todos, followed by their proposed schedule:
\begin{quote}
    ``Student came to the office to discuss schedule for spring 2018 semester, printed and reviewed degree audit, student was shown major map, reviewed EAB for success markers, and notified student. Proposed schedule is: [courses]. Additional comments: [any additional comments].''
\end{quote}
Out of 851 comments left by Advisor 1, 491 begin with ``student came to the office to discuss'' and follow  a structure identical or quite similar to the one above. Interestingly, we notice that Advisor 1 is the only advisor who refers routinely to the use of digital interfaces (``reviewed EAB for success markers'') during advising.

Second, the focus is strictly academic. When Advisors 1 and 2 do personalize their advice to the student, their scope of consideration is also largely constrained to academic topics like placement or exams. For example, Advisor 2 writes: 
\begin{quote}
``Student wants to change major from bio to exercise science... I advised that they could possibly do mini-mester for spring in order to compensate for credits lost towards degree.''
\end{quote}
This difference in structure and focus on academic topics is reflected in the average Jaccard scores and overall distribution of codes.

\subsection{Qualitative Analysis of Advisor Expertise}
\label{appendix:qual_expertise:methods}

\subsubsection{Codebook for Qualitative Analysis of Non-Algorithmic Information}
\label{appendix:qual_deets:qual_interventions}

Below we include our codebook for types of reasons that advisors took interventions.

\begin{itemize}
\item \textbf{Course credits and placement:} Matters to do with course placement and credit, including the student having prior credit from AP/dual enrollment, or discussion of the student's performance on a placement exam for courses.

\item \textbf{Course problem:} The student is enrolled in the wrong course for their skill level, missing a course they should be enrolled in, wants to drop/withdraw from a course they dislike, or is taking a course in which they are performing poorly.

\item \textbf{Degree progress:} The student is behind with degree progression or wants to progress more quickly.

\item \textbf{Documents:} The student has not turned in or checked their financial documents.

\item \textbf{Financial regjstration problem:} The student has a non-payment issue or is on the drop-list due to financial holds/reinstatements.

\item \textbf{Load management:} The student has too light or too heavy a course/extracurricular load.

\item \textbf{Major choice:} The student expresses preferences or indecision about major selection, often intertwined with discussion of career goals.

\item \textbf{Non-financial registration problem:} The student is unable to register due to having various non-financial holds on their account.

\item \textbf{Personal circumstances:} The student has various personal circumstances that are affecting their academic goals (health problems, family hardship, transportation considerations). 

\item \textbf{Low student engagement:} The student did not attend their scheduled appointment, does not have an appointment, or is being standoffish in their current appointment.

\item \textbf{Scholarship process:} The student has a financial aid consideration, such as the student not meeting course requirements for their financial aid, or the student's financial aid not having been posted to their account.

\item \textbf{Student-initiated:} The student explicitly asks the advisor for assistance or advice about an intervention.

\item \textbf{Transfer/dropout:} The student is considering transferring from or dropping out of Georgia State University.

\item \textbf{Unregistered:} The student is not registered for courses for the coming semester.

\end{itemize}

\subsubsection{Comments Mentioning Reasons} 
\label{appendix:qual_results:comments_reasons}
Table \ref{app:table:qual_interventions_summary} records the number of meetings for all interventions (after exclusion of potential transfer students), as well as the number of comments associated with these meetings, the number of times the intervention was mentioned in these meetings, and the number of times the reason was mentioned for taking the intervention.

\begin{table*}[ht]
\centering
 \resizebox{0.9\textwidth}{!}{
\begin{tabular}{l|c|c|c|c|c}
\hline
\textbf{Intervention} & \textbf{Sig.} & \textbf{\# Meetings} & \textbf{\# Comments} & 
\shortstack{\textbf{\# Intervention}\\ \textbf{Mentioned}} & 
\shortstack{\textbf{\# Reason}\\ \textbf{Mentioned}} \\
\hline

\textbf{2} (Correct Schedule) & \checkmark & 756 & 64 & 38 & 36 \\
\textbf{4} (Progression Issues) & \checkmark & 972 & 58 & 27 & 27 \\
\textbf{5} (Course Balance) & $\times$ & 241 & 35 & 15 & 14 \\
\textbf{9} (Fin. Issues) & $\times$ & 626 & 92 & 65 & 65 \\
\textbf{10} (Non-Fin. Requirements) & $\times$ & 348 & 13 & 3 & 3 \\
\textbf{16} (Schedule Appointment) & \checkmark & 1343 & 347 & 254 & 157 \\
\textbf{17} (Recommend Schedule) & \checkmark & 1915 & 791 & 486 & 243$^{*}$ \\
\textbf{20} (Develop Degree Planner) & \checkmark & 2384 & 484 & 12 & 6 \\

\end{tabular}}

\caption{\normalfont Of the significant interventions and the three most insignificant interventions with a substantial number of meetings, counts of how many meetings with that intervention applied have comments left by advisors, comments that actually mention the relevant intervention, and comments that mention the reason(s) for taking that intervention. We can only analyze comments mentioning the reasons for taking an intervention for signs of non-algorithmic, outcome-relevant information. \\
$^*$ Instead of analyzing reasons for taking the intervention for mentions of non-algorithmic information, we analyzed comments for \textit{customizations} of schedules. See Appendix \ref{app:qual_expertise:results:int_17} for  details about schedule customization.\\
}\label{app:table:qual_interventions_summary}

\end{table*}

\subsubsection{Results for Intervention 17 (Recommend Schedule)}
\label{app:qual_expertise:results:int_17}
Without exception, when the advisors mention a reason for recommending a schedule, it is that the student had requested to discuss their schedule. In addition to the 243 mentions of a reason for schedule recommendations, there are 81 instances of advisors customizing recommended schedules to the student. Common customizations include the advisor recommending a specific class to a student based on the course placement test, the advisor suggesting schedule adjustments if the student has a heavy course load or is struggling in a particular class, and the advisor advising a student who is behind on degree requirements to modify their schedule to prioritize required courses and catch up. We note that customizations are likely a vast undercount, as we stringently required explicit evidence linking the information to their schedule recommendations that advisors do not often provide in their comments (for instance, simply mentioning the student wants to be a doctor and listing biology courses would not count; the advisor would have to explicitly say that the courses fulfilled pre-medical requirements).

\subsubsection{Comparison to Insignificant Interventions}
\label{appendix:qual_results:insignificant_ints}

\begin{table}
\centering \resizebox{0.9\textwidth}{!}{
\begin{tabular}{r|r|c|c|c}
\hline
& & \textbf{Intervention 5} & \textbf{Intervention 9} & \textbf{Intervention 10} \\
& & Adjust Course Balance & Address Financial Issue & Address Non-Financial Requirement \\

\hline
1 & Course Credits and Placement & 5 & - & - \\
2 & Course Problem & 6 & - & - \\
3 & Degree Progress & 3 & - & - \\
4 & Documents & - & 5 & - \\
5 & Fin. Registration Problem & - & 52 & - \\
6 & Load Management & 1 & - & - \\
7 & Major Choice & 1 & - & - \\
8 & Non-fin. Registration Problem & - & - & 3 \\
9 & Personal Circumstances & - & 1 & - \\
10 & Scholarship Process & - & 6 & - \\
11 & Student-Initiated & - & 1 & - \\
\hline
\end{tabular}
}
\caption{\normalfont Reasons cited by advisors for taking ExpertTest$^*$ insignificant interventions.}\label{tab:insignificant_interventions}
\end{table}
Results for the insignificant interventions are reported in Table \ref{tab:insignificant_interventions}. Unlike the significant interventions, which incorporate diverse, non-algorithmic information, two of the three insignificant interventions, Intervention 9 (Address Financial Issues) and Intervention 10 (Address Non-financial Collegiate Requirements), are dominated by  administrative holds on registration, most commonly due to nonpayment of a balance on the student's account. These administrative holds are plausibly either predictable from administrative data (for instance, financial holds might be predictable from students' expected financial contribution) or not outcome-relevant (when the holds are resolved).

Intervention 5 (Adjust Course Balance) was an exception; though insignificant, advisors describe  taking the intervention for similar reasons as Intervention 2 (Correct Schedule) and Intervention 4 (Progression Issues). We speculate that the insignificant result might be due to inconsistent use of this label given overlap in its meaning with these interventions. 75\% of the 272 meetings in which advisors indicated applying Intervention 5 (Adjust Course Balance) also indicate applying Intervention 2 (Correct Schedule) or Intervention 4 (Progression Issues), and in our coding, the content of the comments appears quite similar (whether or not intervention 2 or 4 were also checked).

\section{Advisor Heterogeneity Analysis}

\subsection{Heterogeneity Analysis Methods}\label{app:heterogeneity:methods}

To examine heterogeneity in treatment effects by advisor assignment, we estimate linear and logistic regression models predicting graduation outcomes with interaction terms for the most common MAAPS advisors.

\subsubsection{Data Cleaning}
We perform the exclusions described in \ref{sec:data_methods:cleaning}, resulting in a final test population of $n=834$ students ($427$ in treatment and $407$ in control).

We then define the following variables:
\begin{itemize}
    \item \( T_i \in \{0, 1\} \) indicates assignment to the treatment group
    \item \( Z_i \in \{0, 1\} \), indicates that student \( i \) graduated by 2020
    \item  $A_{ij}\in \{0,1\}$ for $j\in[1,3]$, where $A_{ij}=1$ indicating student $i$ was assigned to MAAPS advisor $j$.\footnote{For all students in the control arm, $A_{ij}=0$ for all $j$. When students in the MAAPS treatment arm met with multiple advisors, they are assigned the advisor they met with the most. There are five MAAPS advisors; advisors 4 and 5 are left out of the regression because they did not leave many comments on student meetings, and to avoid multicollinearity.}
    \item $\textsf{Gender}\in \{0,1\}$, with $0$ indicating female and $1$ indicating male.
    \item $\textsf{ACT}_i\in [0,36]$, the student's ACT score. Following the methodology in Rossman et al. \cite{rossmanMAAPSAdvisingExperiment2023}, we convert composite SAT scores to ACT scores using a concordance table provided by the Educational Testing Service and ACT Inc. \cite{ACTSATConcordanceTables2016}.
    \item $\textsf{EFC}_i\in [1,14]$ a scale indicating the level of the student's expected financial contribution (from $\$0 - \$1,050$ at level $0$ to $\$22,051+$ at level $14$).
    \item $\textsf{Hours}_i\in \mathbb{Z}$ indicating the number of college credit hours earned by the student before enrollment at GSU, including AP or IB credit.
\end{itemize}

\subsubsection{Regression}

To examine heterogeneity in treatment effects by advisor assignment, we estimate a linear regression model predicting the binary graduation outcome:

\begin{align*}
\label{eq:advisor_logreg}
    \mathbb{P}(Z_i = 1) = & 
    \beta_0 + \beta_1 T_i + \sum_j \theta_j A_{ij} + \gamma_1 \textsf{Gender}_i  + \gamma_2 \textsf{ACT}_i + \gamma_3 \textsf{EFC}_i  + \gamma_4 \textsf{Hours}_i
\end{align*}

We estimate an analogous logistic regression model to ensure consistency of advisor coefficient rankings.

\subsection{Heterogeneity Analysis Results}\label{app:heterogeneity:results}

While we do not find significant heterogeneous treatment effects across advisors, both linear (Table \ref{tab:ols_advisor_htes}) and logistic regression (Table \ref{tab:logistic_advisor_htes}) produce consistent ranking of advisor-specific coefficients --- with Advisor $3$ having the highest coefficient ($0.0685$ in the linear regression), followed by Advisor 2, and last Advisor 1.

\begin{table*}[h!]
    \centering
\begin{tabular}{lccccc}
        \hline
        \multicolumn{6}{c}{\textbf{OLS Regression Results: Advisor Heterogeneity in Graduation}} \\
        \hline
        Variable & Coefficient & Standard error & $t$-statistic & $p$-value & 95\% CI \\
        \hline
        $\beta_0$ & 0.2857 & 0.024 & 11.945 & 0.000 & [0.239, 0.333] \\
        $T_i$ & 0.0573 & 0.082 & 0.700 & 0.484 & [-0.103, 0.218] \\
        $A_i^1$ & -0.0833 & 0.088 & -0.945 & 0.345 & [-0.256, 0.090] \\
        $A_i^2$ & -0.0125 & 0.087 & -0.144 & 0.886 & [-0.183, 0.158] \\
        $A_i^3$ & 0.0685 & 0.087 & 0.785 & 0.433 & [-0.103, 0.240] \\
        $\textsf{Gender}$ & -0.1429 & 0.032 & -4.524 & 0.000 & [-0.205, -0.081] \\
        $\textsf{ACT}$ & 0.0456 & 0.016 & 2.811 & 0.005 & [0.014, 0.078]\\
        $\textsf{EFC}$ & -0.0082 & 0.015 & -0.550 & 0.583 & [-0.038, 0.021] \\
        $\textsf{Hours Other}$ & 0.0200 & 0.016 & 1.259 & 0.208 & [-0.011, 0.051] \\
        \hline
        \multicolumn{6}{c}{Observations: 834, Adjusted R-squared: 0.036, F-statistic: 4.889, F-statistic $p$-value: $6.48\times 10^{-6}$ } \\
        \hline
    \end{tabular}
    \caption{\normalfont OLS regression estimating the effect of student assignment to a particular advisor on graduation outcomes. The model includes controls for gender, ACT scores, expected family contribution (EFC), and other academic hours applied toward the degree. Coefficient estimates, associated $t$-statistics, $p$-values, and 95\% confidence intervals are reported.}\label{tab:ols_advisor_htes}
\end{table*}

\begin{table*}[h!]
    \centering
    \begin{tabular}{lccccc}
        \hline
        \multicolumn{6}{c}{\textbf{Logistic Regression Results: Advisor Heterogeneity in Graduation}} \\
        \hline
        Variable & Coefficient & Standard error & $z$-statistic & $p$-value & 95\% CI \\
        \hline
        $\beta_0$ & -0.9454 & 0.127 & -7.419 & 0.000 & [-1.195, -0.696] \\
        $T_i$ & 0.3265 & 0.438 & 0.745 & 0.456 & [-0.533, 1.186] \\
        $A_i^1$ & -0.4799 & 0.479 & -1.002 & 0.316 & [-1.419, 0.459] \\
        $A_i^2$ & -0.0877 & 0.465 & -0.188 & 0.851 & [-0.999, 0.824] \\
        $A_i^3$ & 0.3160 & 0.460 & 0.687 & 0.492 & [-0.586, 1.218] \\
        $\textsf{Gender}$ & -0.8024 & 0.182 & -4.404 & 0.000 & [-1.160, -0.445] \\
        $\textsf{ACT}$ & 0.2466 & 0.088 & 2.879 & 0.005 & [0.073, 0.420]\\
        $\textsf{EFC}$ & -0.0407 & 0.082 & -0.496 & 0.620 & [-0.202, 0.120] \\
        $\textsf{Hours Other}$ & 0.0902 & 0.082 & 1.098 & 0.272 & [-0.071, 0.251] \\
        \hline
        \multicolumn{6}{c}{Observations: 834, Log likelihood: -458.97, Log likelihood null: -477.05, LLR $p$-value: $7.044\times 10^{-6}$ } \\
        \hline
    \end{tabular}
    \caption{\normalfont Logistic regression estimating the effect of student assignment to a particular advisor on graduation outcomes. The model includes controls for gender, ACT scores, expected family contribution (EFC), and other academic hours applied toward the degree. Coefficient estimates, associated $z$-statistics, $p$-values, and 95\% confidence intervals are reported.}
    \label{tab:logistic_advisor_htes}
\end{table*}

\end{document}